\newcommand{\tsecompldate}{13th April 2001}

\documentclass[12pt,a4paper]{article}

\usepackage{amssymb}

\newcommand{\vol}[1]{{\bf #1}}

\newcommand{\tpaptitle}[1]{}
\newcommand{\tpretitle}[1]{}
\newcommand{\tarttitle}[1]{}
\newcommand{\tproctitle}[1]{in ``#1''}
\newcommand{\tbktitle}[1]{``#1''}
\newcommand{\tpreprint}[1]{Report No. #1}
\newcommand{\tpreeprint}[1]{}
\newcommand{\tISBN}[1]{}

\newcommand{\tref}[1]{(\ref{#1})}

\newcommand{\tnotpre}[1]{#1}
\newcommand{\tversion}{Final version}

\newcommand{\tpre}[1]{}
\newcommand{\tprenote}[1]{}

\newcommand{\tnote}[1]{}
\newcommand{\tcomment}[1]{}

\newcommand{\href}[2]{#2}
\newcommand{\eprint}[1]{{\tt #1}}

\newcommand{\tsedevelop}[1]{{}}

\newcommand{\tsevec}[1]{\vec{#1}}
\newcommand{\tsemat}[1]{{\mathbf #1}}
\newcommand{\tsebb}[1]{\mathbb{#1}}

\makeatletter

\@addtoreset{equation}{section} \makeatother \renewcommand{\tversion}{Preprint version}
\renewcommand{\tnotpre}[1]{}
\renewcommand{\tpre}[1]{#1}
\renewcommand{\tprenote}[1]{\footnote{#1}}
\renewcommand{\tpaptitle}[1]{{\it #1},}
\renewcommand{\tpretitle}[1]{{\it #1},}
\renewcommand{\tarttitle}[1]{{\it #1},}
\renewcommand{\tproctitle}[1]{in ``#1'',}
\renewcommand{\tbktitle}[1]{``#1''}
\renewcommand{\tpreprint}[1]{preprint no.\ #1}
\renewcommand{\tpreeprint}[1]{ [\eprint{#1}]}
\renewcommand{\tISBN}[1]{ (ISBN {\tt #1})}
\newcommand{\half}{\frac{1}{2}}
\newcommand{\bea}{\begin{eqnarray}}
\newcommand{\eea}{\end{eqnarray}}
\newcommand{\beq}{\begin{equation}}
\newcommand{\eeq}{\end{equation}}
\newcommand{\nnel}{\nonumber \\ {}}

\newcommand{\veck}{\tsevec{k}}

\newcommand{\vecx}{\tsevec{x}}
\newcommand{\phivec}{\tsevec{\phi}}
\newcommand{\Psivec}{\tsevec{\Psi}}

\newcommand{\Dmat}{\tsemat{D}}
\newcommand{\Gmat}{\tsemat{G}}

\newcommand{\bbZ}{\tsebb{Z}}

\newcommand{\calF}{{\cal F}}
\newcommand{\calL}{{\cal L}}

\newcommand{\ahat}{\widehat{a}}
\newcommand{\bhat}{\widehat{b}}
\newcommand{\Hhat}{\widehat{H}}
\newcommand{\Qhat}{\widehat{Q}}

\newcommand{\kbar}{\bar{k}}

\newcommand{\ahatk}{\ahat_{\veck}}
\newcommand{\bhatk}{\bhat_{\veck}}
\newcommand{\omegak}{\omega_{\veck}}

\newcommand{\unitmat}{\tsemat{\hbox{{\bf 1}\kern-.28em{\bf I}}}}
\newcommand{\dslash}{{d\kern-.45em{/}}}
\newcommand{\deltaslash}{{\delta\kern-.28em{'}}}

\newcommand{\dkf}{ \frac{dk_4}{(2\pi)^{1}} }
\newcommand{\dfk}{ \frac{d^4k}{(2\pi)^{4}} }
\newcommand{\dfek}{ \dslash^{4-2\epsilon}k }
\newcommand{\intdtk}{ \int \frac{d^3 \veck}{(2\pi)^{3}} }
\newcommand{\intdfk}{ \int \dfk }

\newcommand{\tr}{{\rm tr}}
\newcommand{\Tr}{{\rm Tr}}
\newcommand{\TR}{{\rm TR}}
\newcommand{\DET}{{\rm DET}}

\newcommand{\lvac}{\langle 0 |}
\newcommand{\rvac}{| 0 \rangle}

\begin{document}

\renewcommand{\thefootnote}{\fnsymbol{footnote}}

 \tpre{\begin{flushright}
 {\tt Imperial/TP/0-01/17} \\
 \eprint{hep-th/0104128}
 \\
 \tsecompldate \\
 \tsedevelop{({\tt fbg.tex} \tversion LaTeX-ed on \today ) \\}
 \end{flushright}
 \vspace*{1cm} }

\begin{center}
{\Large\bf Vacuum Energy Densities and \\ Multiplicative Anomalies}\\
 {\Large\bf in a Free Bose Gas}\\
 \tpre{\vspace*{1cm} }
 {\large T.S. Evans\footnote{email:
\href{mailto:T.Evans@ic.ac.uk}{{\tt T.Evans@ic.ac.uk}}\tnotpre{,
\tsecompldate}\tpre{, WWW:
\href{http://theory.ic.ac.uk/links/time}{\tt
http://theory.ic.ac.uk/\symbol{126}time} }}}
 \\
 \tpre{\vspace*{1cm}}
 \href{http://theory.ic.ac.uk/}{Theoretical Physics},
 Blackett Laboratory, Imperial College,\\
 Prince Consort Road, London, SW7 2BW,  U.K.
 \tnotpre{\\
 Tel: U.K.-20-7594-7837,
 Fax: U.K.-20-7594-7844 (or -7777) 
 }
\end{center}

\begin{abstract}

The vacuum energy density or free energy of a free charged Bose
gas at non-zero densities is studied in the context of the debate
about Multiplicative Anomalies.  Some $\zeta$-function regularised
calculations of the free energy in the literature are reexamined,
clarified and extended.  A range of apparently distinct answers
can obtained.  Equivalent dimensional regularisation results are
also presented for comparison. I conclude that operator ordering
and normal ordering are not responsible for these differences.
Rather it is an undesirable but unavoidable property of
$\zeta$-function regularisation which leads to these different
results, making it a bad scheme in general. By comparison I show
how dimensional regularisation calculations give a consistent
result without any complications, making this a good scheme in
this context.

\end{abstract}


{}\tnote{tnotes such as this not present in final version}

\renewcommand{\thefootnote}{\arabic{footnote}}
\setcounter{footnote}{0}

\section{Introduction}

The vacuum energy density in QFT (quantum field theory) is a
non-trivial object which leads to interesting physical phenomena
such as inflation or the Casimir effect.  When using path
integrals to calculate it, one often encounters terms of the form
$ \ln\det ( \Delta^{-1} ) $ where $\Delta$ is the propagator for
some field.  However, in QFT this is the determinant of an
infinite dimensional matrix, and this is usually infinite. It is
therefore extremely interesting to find that the ``Multiplicative
Anomaly'' $a(A,B)$
\beq
 a(A,B) :=  \ln\det ( AB ) - \ln\det (A) - \ln\det ( B ),
 \label{adef}
\eeq
need {\em not} be zero for two pseudo-differential operators $A$
and $B$, \cite{Kass,EVZ,KV}.  The term anomaly in this context is
used to indicate the failure of a familiar algebraic expression
rather than a breakdown of a classical symmetry in the quantum
theory. Since the individual terms in \tref{adef} are usually
naively infinite in QFT, one must regulate before any serious
discussion. A subscript on a quantity will be used to indicate the
regularisation scheme used e.g.\ $a_R, [\ln\det(A)]_R$ for some
scheme R. The first examples of a non-zero Multiplicative Anomaly
in QFT was given by Elizalde, Vanzo and Zerbini \cite{EVZ} using
$\zeta$-function regularisation of Dowker and Critchley \cite{DC}
and two free scalar fields in four or more even dimensional
Euclidean space time. The Multiplicative Anomaly $a_\zeta$ of
\tref{adef} is related to the Wodzicki residue \cite{Wo,Kass}
which is important for non-commutative geometry
\cite{Connes}.\tprenote{See Connes \cite{Connes}, page 307.}

There have been two types of extension to the original work of
\cite{EVZ}. Firstly, one can investigate different renormalisation
schemes. In an earlier paper I showed that loop momentum cutoffs,
including space-time lattices and dimensional regularisation ---
schemes common in particle physics --- are all free of this
Multiplicative Anomaly \cite{TSEanom}.  Rather, non-zero
Multiplicative Anomalies are a feature of the whole family of
Schwinger proper time regularisation schemes of which
$\zeta$-function methods are just one example \cite{Ball}. The
second direction has been to show that the original Multiplicative
Anomaly of \tref{adef} is just one of many algebraic identities
which $\zeta$-function regularisation expressions fail to obey
\cite{TSEanom,EFVZ,MT98,MT,FSZ,Su,CZ,Fi}.  If one is using
$\zeta$-function regularisation, the importance of these
Multiplicative Anomalies is not in doubt from the point of view of
mathematical consistency.  Furthermore there are several types of
problem where $\zeta$-function regularisation methods are probably
the best method to use, for example Casimir effect problems or
problems of QFT in curved space time. The important debate is
whether or not the extra terms due to Multiplicative Anomalies
lead to new physics.  A more extensive summary of Multiplicative
Anomalies is given in appendix \ref{appma}.

One of the best examples to study in the discussion of physical
meaning of Multiplicative Anomalies is that of the free Bose gas.
This is because even in a free theory with one complex
relativistic field, the U(1) symmetry can be broken and
Bose-Einstein condensation then takes place \cite{Ka}. In this
model, the calculations of Elizalde et al \cite{EFVZ} (see also
\cite{Fi}) of the free energy, using $\zeta$-function
regularisation methods, gave a result which has extra terms when
compared to standard results found in texts such as Kapusta
\cite{Ka}. The extra terms of \cite{EFVZ} were chemical potential
but not temperature dependent, and they led to alterations of the
critical temperature. Furthermore in \cite{EFVZ} these extra
terms, which come from the Multiplicative Anomaly, were shown to
be essential for mathematical consistency. Elizalde et al.\
suggested that normal ordering might be relevant in understanding
why these extra terms appear.  McKenzie-Smith and Toms
\cite{MT98,MT} then looked at the free Bose gas in
$\zeta$-function regularisation with particular reference to
canonical calculations and normal ordering rather than path
integral calculations alone. They concluded that there is no
problem if one follows canonical definitions.

Unfortunately there are mathematical inconsistencies between some
of these $\zeta$-function regularisation results for the free Bose
gas.  I will reexamine the claims and calculations of Elizalde et
al.\ \cite{EFVZ}, and McKenzie-Smith and Toms \cite{MT98,MT}
within $\zeta$-function regularisation, and will show how to fit
them into a consistent mathematical framework.  However, I confirm
the mathematical results of \cite{EFVZ} which show that one can
reasonably obtain a number of different results for the free
energy of the free Bose gas when using $\zeta$-function
regularisation. However I differ from previous authors in my
interpretation of the physics contained in these results.  I
conclude that canonical methods, and normal ordering in
particular, offer no explanation for the genuinely different
answers obtained. All of the many formal starting points are
equally good. Rather my explanation is that in certain
regularisation schemes, such as $\zeta$-function regularisation,
the UV divergences are controlled by functions whose form depends
on physical parameters such as chemical potentials. Equivalently
the renormalisation scale in $\zeta$-function regularisation
depends on some of the physical parameters of the problem. Hence
the comparison of results calculated in $\zeta$-function
regularisation at different values of physical parameters is then
extremely difficult.  I conclude that there is no new physics of
the free Bose gas in the non-standard results provided by some
$\zeta$-function regularisation calculations.

However, in order to discuss non-standard results, I must also
construct a standard.  Thus to compare the $\zeta$-function
regularisation results for the free Bose gas of
\cite{EFVZ,MT98,MT} with the free energy results found in the
literature such as \cite{Ka}, it is convenient to use a standard
particle physics regularisation scheme. Dimensional
regularisation, as defined in the particle physics
literature\footnote{Confusingly, some of the $\zeta$-function
regularisation literature, such as \cite{CZ}, refer to some
Schwinger proper-time regularisation schemes as ``dimensional
regularisation''.  This is not the same as the scheme widely used
in particle physics, as noted by Ball \cite{Ball} and discussed in
appendix \ref{appzfr} around equation \tref{SPTDRdef}.}
\cite{Co,PDG}, is used here for several reasons. Firstly it is
widely used in practical calculations \cite{PDG}. Secondly it has
no Multiplicative Anomalies when used in a sensible and correct
manner \cite{TSEanom}, and so is in the same position as several
other standard schemes such as lattice regularisation and simple
momentum cutoffs\tnote{In both cases though perhaps there are
problems if I regulate in the time direction?}.  Lastly,
dimensional regularisation and $\zeta$-function regularisation
regulate by inserting a non-integer power of a polynomial of loop
momentum, so one often finds that the same standard integral is
required for both schemes. Therefore I will discuss the relation
of $\zeta$-function regularisation calculations of free Bose gas
to the standard results obtained with dimensional regularisation
and in doing so make contact with results found in the literature
such as are given in \cite{Ka}.

In the next section I examine how various formal, and therefore
strictly meaningless, expressions for the free Bose gas are
derived.   I give precise well defined mathematical expressions
only in the following section by carefully defining the
regularisations.  In section 4 I study the relationships between
the precise and well defined forms of the free energy.  Section 5
is devoted to looking at the physical content of the expressions
and my conclusions are given in the final section.  I have tried
to put as many technical details in the first appendix, but
because the confusion in the literature I feel it is important to
specify the precise mathematical approach used here.  Appendix B
puts the multiplicative anomaly found in the free Bose gas in a
wider context.

\section{Formal expressions for the free energy}\label{formaldef}

The free Bose gas is a heat bath of charged non-interacting scalar
particles.  The dynamics are described by the usual Klein-Gordon
Lagrangian for a free relativistic complex scalar
field\footnote{The condensed matter limit was considered
separately in \cite{MT98}.  It is contained in this analysis as
the $m \sim \mu \gg T \sim m-\mu \geq 0$ limit \cite{TSEdal}
though regularisation prescriptions may differ from that used in
\cite{MT98}.}
\begin{equation}\label{Lagdef}
  {\cal L} = |\partial_\mu \Phi|^2 - m^2 |\Phi|^2 .
\end{equation}
For simplicity I work in four-dimensional Euclidean space, of
which the spatial dimensions have volume $V$, as results such as
\cite{EVZ} show that Multiplicative Anomalies are often
non-trivial in such space-times. The Lagrangian \tref{Lagdef} is
invariant under a global phase transformation which is associated
with the conservation of particle minus anti-particle
number\footnote{The number of particles and the number of
anti-particles in each mode is also separately conserved but these
symmetries are lost in interacting theories so I do not consider
them here.}. To describe the statistical average I use the density
matrix of an equilibrium grand canonical ensemble, for which the
temperature and the chemical potential of the heat bath (intensive
variables) are specified rather than the energy and charge
(extensive variables). The Euclidean approach to thermal field
theory encodes this very simply and this is sufficient for our
purposes. I will use the approach in which the temporal direction
is made periodic with length $\beta$ while the effects of non-zero
chemical potential are included by working with an effective
Hamiltonian, corresponding to the Lagrangian density\footnote{The
alternative Euclidean method puts both temperature and chemical
potential in the boundary conditions \cite{LvW,TSEdal}.}\tnote{I
have also considered some of the calculations using this
alternative approach but it appears to give the same answers.}
\cite{Ka,LvW,TSEdal}
\begin{equation}\label{Leff}
 {\cal L}_\mu =  \Phi^\dagger K_+ \Phi ,
\end{equation}
where
\begin{eqnarray} \label{Kdef}
  K_\pm &:=& (\partial_4 \pm \mu)^2 + \omega^2
  \\
  \label{omegadef}
  \omega^2 &:=& -\tsevec{\nabla}^2 + m^2  .
\end{eqnarray}
The $\tsevec{\nabla}$ acts only on the three spatial directions.
The Euclidean temporal derivatives, $\partial_4$, are just shifted
by the chemical potential $\mu$ (a real parameter) corresponding
to a real shift in the origin of Minkowski energies
\cite{TSEdal}\footnote{The meaning of Hermitian conjugation has
been modified in the appropriate manner for Euclidean theories.}.
Consider the partition function $Z$ and the associated free energy
density $F$
\begin{equation}\label{ZFdef}
  Z := \Tr \{ e^{-\beta(H-\mu Q) } \} , \; \; \;
  F := - \frac{1}{\beta V} \ln (Z) .
\end{equation}
I will keep $|\mu| < m$ which, in a free theory, means working in
the symmetric phase \cite{Ka}. Likewise, I keep $T>0$ as at zero
temperature all particles will be in the ground state so $|\mu|=0$
or $m$. Since the theory is free, one may quickly obtain an
expression for $F$ but it is crucial to examine the familiar
steps, given that the failure of supposedly familiar algebraic
identities such as \tref{adef} is at the centre of the
Multiplicative Anomaly debate.

Before I look at these calculations, note immediately that in this
section I am writing {\em formal} expressions, by which I mean
they have not yet been regularised, so they are na\"{\i}vely
infinite and therefore strictly meaningless.  In doing this I am
merely reflecting standard QFT procedures and will correct this
later sections.  This distinction, between formal infinite
expressions and their meaningful regularised counterparts, is
however not always clearly maintained in the literature.  One of
the aims of this paper is to indicate clearly when one is
performing formal manipulations, and when rigorous manipulations
of finite objects is being performed.

\subsection{Path integral approach}

When working with the path integral in this model, one would
normally exploit the fact that the field is complex and work in
the one-dimensional complex irreducible representation of U(1).
The path integral for $Z$ is
\begin{equation}\label{Zcompdef}
  Z = \int D\Phi D\Phi^\dagger \exp \{ -\int d^4x \; \calL_\mu[\Phi,\Phi^\dagger] \}
\end{equation}
so the usual results for complex Gaussian integration
give\footnote{One often sees the $\ln \det$ notation in
$\zeta$-function regularisation work rather than $\Tr \ln$.
However, neither has any proper definition unless regularised, and
I am free to choose an appropriate definition, {\it pace}
\cite{EFVZ3}. I will define them to be equal to the same
regularised expression. In any case, all practical regularised
expressions I know involve a sum or integral, i.e.\ they look like
a trace and do not contain products which might remind one of a
finite determinant.  After all, the most familiar definition of a
Riemann $\zeta$-function involves a sum over positive integers,
rather than the alternative formula of a product over prime
numbers. I will therefore tend to use the $\Tr \ln$ notation,
though I treat the two as equivalent.}\tnote{This means that even
in the $T>0$ terms they are even in $\mu$ so I don't need to
calculate both terms.  Why? Is it obvious? Is it true?}
\begin{eqnarray}
 \label{F4Adef0}
 F_{4K+} & := & - \frac{1}{\beta V} \ln \det (K_+)
 = - \frac{1}{\beta V} \Tr \ln (K_+)  .
\end{eqnarray}
Here the trace $\Tr$ with a capital T and lower case r indicates
that the trace is over the four dimensional Euclidean space so
that in coordinate representation
\begin{equation}\label{Trdef}
  \Tr \equiv \int_0^\beta dt_4 \int_V d^3 \vecx
\end{equation}

If I work with the reducible complex two dimensional
representation of U(1), i.e.\ use a vector $\Psivec =(\Phi,
\Phi^\dagger)$, then the Lagrangian is given by
\begin{equation}\label{L2def}
   {\cal L}_\mu =  \half \Psivec^\dagger \Dmat^{-1} \Psivec
\end{equation}
where
\begin{equation}\label{Ddef}
  \Dmat^{-1} =
  \left[ \begin{array}{cc}
  K_+& 0 \\
  0 &  K_-
  \end{array} \right].
\end{equation}
The path integral for $Z$ is
\begin{equation}\label{Zvecdef}
  Z = \int D\Psivec \; \exp \{ -\int d^4x \; \calL_\mu [\Psivec] \}
\end{equation}
and Gaussian integration gives
\begin{eqnarray}
 \label{F4Adef0b}
 F_{4A} & := & - \half  \ln \DET (\Dmat^{-1})
  =  - \half  \TR \ln (\Dmat^{-1})
\end{eqnarray}
where $\TR$ and $\DET$ are now taken over both the two-dimensional
field space and the infinite dimensions coming from the
four-dimensional space-time. One can take the determinant over the
two-dimensional field space \cite{EFVZ2} to give
\begin{eqnarray}
 \label{F4Adef}
 F_{4A} & := & - \half  \Tr \ln (A)
\end{eqnarray}
where
\begin{equation}\label{Adef}
  A = K_+ . K_-  .
\end{equation}

One often sees this calculation expressed in terms of two real
fields, the real and imaginary parts of the complex field $\Phi=
(\phi_1 + i \phi_2)/\surd 2$, e.g.\ \cite{Ka}.  Writing in a real
vector notation $\phivec = (\phi_1,\phi_2)$ gives
\begin{equation}\label{L2defb}
   {\cal L}_\mu =  \half \phivec^T \Gmat^{-1} \phivec
\end{equation}
where
\begin{equation}\label{Ddefb}
  \Gmat^{-1} =
  \left[ \begin{array}{cc}
  -\partial_4^2  - \nabla^2 + m^2 - \mu^2 & \mu \partial_4  \\
  - \mu \partial_4   &  -\partial_4 - \nabla^2 + m^2  - \mu^2
  \end{array} \right].
\end{equation}
The path integral for $Z$ is
\begin{equation}\label{Zrvecdef}
  Z = \int D\phivec \; \exp \{ -\int d^4x \; \calL_\mu [\phivec]
  \}  .
\end{equation}
In this case the propagator has off diagonal elements proportional
to $\mu$, but the eigenvalues are the diagonal entries of $\Dmat$
in \tref{Ddef} and the eigenvectors are $\Phi$ and
$\Phi^\dagger$. These complex fields are of course the U(1) charge
eigenstates and are therefore the most appropriate basis for
unbroken symmetry problems. In any case one obtains the same
result as before \tref{F4Adef}.

One can now look at a strange variation of the calculation of
$F_{4A}$.  One can take \tref{Zrvecdef} and formally one can
factorize it into two integrations
\begin{equation}\label{Zrfactdef}
  Z = Z_+ . Z_-, \; \; \;
  Z_\pm = \int D\Phi D\Phi^\dagger \exp \{ -\int d^4x \half \Phi^\dagger K_\pm \Phi \}
\end{equation}
to give the formula
\begin{eqnarray}
\label{F4Kdef}
  F_{4K} &:=& - \frac{1}{\beta V} \half \Tr \ln (K_+)
  - \frac{1}{\beta V} \half \Tr \ln (K_-)
\end{eqnarray}
Now I see that this seemingly innocent variation with an initial
factorization in the path integral, is producing a result which is
nothing more than a factorization of the quartic operator $A$ of
the expression $F_{4A}$ \tref{F4Adef} into two quadratic
operators. However, it is exactly the success or failure of this
sort of factorization which is measured by the Multiplicative
Anomaly \tref{adef}, i.e.\
\bea
a(K_+,K_-) &=& {2\beta V}\left( F_{4K} -  F_{4A}\right) .
 \label{F4AdefK}
\eea
Thus a non-zero Multiplicative Anomaly says that I can not do
these trivial maneuvers in the path integral such as described
here. Since so much work is based on such formal manipulations of
the path integral, Multiplicative Anomalies have potentially very
grave implications.

Returning to the formal expressions for the free Bose gas free
energy, note that there are many other factorizations of the
operator $A$.  I will follow \cite{EFVZ} and consider just one
other, namely
\begin{eqnarray}
\label{Ldef}
  L_\pm &:=& -\partial_4^2 + (\omega \pm \mu)^2
\end{eqnarray}
where $\omega$ was given in \tref{omegadef}. I could therefore
equally well define the free energy density as
\begin{eqnarray}
\label{F4Ldef}
  F_{4L} &:=& - \frac{1}{\beta V} \half \Tr \ln (L_+)
  - \frac{1}{\beta V} \half \Tr \ln (L_-)
\end{eqnarray}
Again, the Multiplicative Anomaly expresses the failure of
algebraic identities needed to relate $F_{4A}$ to $F_{4l}$ and we
have from \tref{adef}
\bea
a(L_+,L_-) &=& {2\beta V}\left( F_{4L} -  F_{4A}\right)
 \label{F4AdefL}
\eea
This shows again that the Multiplicative Anomalies encode
differences between formally equivalent definitions of the free
energy of a free Bose gas.

Several other combinations also exist but I will focus on just
the $L$ and $K$ factorizations. One common feature of all these
expressions is that they involve four dimensional traces,
equivalently are evaluated using integrals over energy and
momentum, and that they are most naturally obtained within a path
integral approach.

\subsection{Canonical approach}

The free Bose gas gas is also easily obtained using canonical
methods and, as emphasized by McKenzie-Smith  and Toms
\cite{MT98,MT}, this leads to additional familiar forms. I will
start with the Hamiltonian and charge operators for a relativistic
free Bose gas. Unlike the path integral with its c-numbered
fields, I must specify an operator ordering. My first definition
of a Hamiltonian and charge operator are ones which have been
normal ordered in the conventional zero temperature
sense\footnote{Interestingly, at $T>0$ or $\mu \neq 0$ the most
appropriate normal ordering is different and depends on $T$ and
$\mu$ explicitly \cite{ES}.}, i.e.\ annihilation operators to the
right, which I will denote with a subscript $N$, namely
\begin{eqnarray}\label{HNdef}
\Hhat_N &:=& \intdtk \left( \omegak  \ahatk ^\dagger \ahatk
+\omegak  \bhatk ^\dagger \bhatk  \right) ,
\\
\label{QNdef} \Qhat_N &:=& \intdtk \left( \ahatk ^\dagger \ahatk -
\bhatk ^\dagger \bhatk   \right) ,
\end{eqnarray}
with dispersion relation $\omegak$. Right at the start note that
I am working on mass shell with integrals over three momenta, not
traces over four space-time coordinates. Thus the expressions
obtained will have a trace over the three dimensional spatial
coordinates, denoted with lower case letters, as
\begin{equation}\label{trdef}
  \tr \equiv \int_V d^3\vecx ,
\end{equation}
to distinguish it from the full four-dimensional trace $\Tr$ of
\tref{Trdef}.

The vacuum state of the Fock space associated with the $\ahatk $
and $\bhatk $ operators is
\begin{equation}\label{vacdef}
 \ahatk  \rvac = \bhatk  \rvac =0
\end{equation}
This vacuum state will be the true physical vacuum provided we
avoid symmetry breaking, i.e.\ if I choose $m^2>0$ and if I have
low charge densities so that $|\mu| < m$ so there is no
Bose-Einstein condensation.

With these definitions, I see that the expectation values of both
$\Hhat_N$ and $\Qhat_N$ in the vacuum state are zero,
\begin{equation}\label{HQNexp}
 \lvac \Hhat_N \rvac =
 \lvac \Qhat_N \rvac = 0  .
\end{equation}
The density matrix in terms of these normal ordered operators is
just
\begin{equation}\label{rhondef}
  \widehat{\rho}_N
  := \exp \{ -\beta (\Hhat_N - \mu \Qhat_N )\}
  = \exp \{ - \beta \intdtk
  \intdtk \left( (\omegak  - \mu ) \ahatk ^\dagger \ahatk
 +(\omegak  + \mu)\bhatk ^\dagger \bhatk   \right)
 .
\end{equation}

Another ordering is the symmetric one, which I will denote with a
subscript $S$, namely
\begin{eqnarray}\label{HSdef}
\Hhat_S &:=& \intdtk
 \left( \half \omegak  ( \ahatk ^\dagger  \ahatk +\ahatk  \ahatk ^\dagger )
 + \half \omegak  (\bhatk ^\dagger \bhatk  + \bhatk  \bhatk ^\dagger )
 \right),
\\
\label{QSdef}
 \Qhat_S &:=& \intdtk
 \left( \half (  \ahatk ^\dagger \ahatk
               + \ahatk  \ahatk ^\dagger )
 - \half (  \bhatk ^\dagger \bhatk
          + \bhatk  \bhatk ^\dagger ) \right) .
\end{eqnarray}
Such a symmetric operator ordering is often the ordering found
necessary to match path integral results, though a path integral
is written in terms of commuting objects.\tnote{Can I find a
reference for this?} Note that, at least formally, the expectation
value of this Hamiltonian in the vacuum state is now not zero, but
is equal to a half $\omegak $ per mode in the system. On the other
hand the expectation value of this charge operator is still
formally zero,
\begin{equation}\label{HQSexp}
\lvac \Hhat_N \rvac = 2 \left( \intdtk \half \omegak  \right), \;
\; \; \lvac \Qhat_N \rvac = 0  .
\end{equation}
The symmetric ordering density matrix is then just
\begin{eqnarray}\label{rhosdef}
  \widehat{\rho}_S &:=& \exp \{ -\beta (\Hhat_S - \mu \Qhat_S )\}
  \\
  &=& \exp \{ - \beta \intdtk
  \left( \half (\omegak - \mu) ( \ahatk ^\dagger
     \ahatk +\ahatk  \ahatk ^\dagger )
  + \half (\omegak + \mu)
    (\bhatk ^\dagger \bhatk
     + \bhatk  \bhatk ^\dagger )
  \right)
   .
\end{eqnarray}

One can imagine other operator orderings.  In particular, the only
way to get charge operators with a non-zero vacuum expectation
value is to change the ordering from that of $\Qhat_N$ by
different amounts for the $\ahatk $ and $\bhatk $ operators to
give a charge operator equal to $\Qhat_{AS} = \Qhat_N + \alpha$
where $\alpha$ is a c-number. This is equivalent to changing the
definition of zero of charge.  It will not effect the physics as
all such $\Qhat_{AS}$ operators still commute with the
Hamiltonian, but the labels I give to a given physical state will
depend on such decisions. Each (anti-)particle still carries
charge $+1$ ($-1$) but now the state with equal numbers of
particles and anti-particles has charge $\alpha$.  This does not
seem useful physically and further I shall show that normal
ordering has nothing to do with the anomalous terms of interest
here.  It is therefore sufficient to focus on just the normal
ordered \tref{HNdef} and symmetric \tref{HSdef} versions.

The canonical calculation runs as follows.  Since it is a free
theory, the Hilbert space is a direct product of the single mode
states, that is a direct product of the Fock spaces associated
with the single oscillators $\ahatk (\veck)$ or $\bhatk (\veck)$
for all possible $\veck$. In practice this means that for the
symmetric density matrix \tref{rhosdef}
\bea
 Z_S &=& \Tr \{ e^{ -\beta (\Hhat_S - \mu \Qhat_S )} \}
 \\
 &=& \left( \prod_{\veck} {\tr}_{a,\veck} \{
 \exp \{ -\beta V (\omegak  - \mu)
          \ahatk ^\dagger \ahatk  \} \} \right)
          \nnel
 &&
 \; \; \; \; \; \; \;
 \times
 \left( \prod_{\veck} {\tr}_{b,\veck} \{
 \exp \{ -\beta V (\omegak  + \mu)
          \bhatk ^\dagger \bhatk  \} \} \right)
\eea
where here the traces, $\tr$, are over the appropriate single
oscillator Fock spaces.   The direct product of the Hilbert space
has resulted in a factorisation of terms in this formal
expression.  Standard calculations then give\tnote{Needs some more
details to show factorization is going on.}
\begin{eqnarray}
 F_{3\mu} &:=&
 \sum_\pm \left[ \half \tr \{ \omegak \pm \mu \}
 + \half \tr \{\ln [ 1-\exp (-\beta(\omegak \pm \mu))]\} \right] .
 \label{F3mudef}
\end{eqnarray}
In exactly the same way, the normal ordered form for the density
matrix \tref{rhondef} gives a second formal expression
\begin{eqnarray}
 F_{3N} &:=&
 \sum_\pm \left[ \half \tr \{\ln [ 1-\exp (-\beta(\omegak \pm \mu))]\} \right]
 \label{F3Ndef}
\end{eqnarray}
with the $\beta$ independent zero-point energy term missing as one
would expect.

However,  there is yet another formal expression discussed in the
literature.  This I will call $F_3$ where
\begin{eqnarray}
 F_{3} &:=&
 \tr \{ \omegak   \}
 + \half \tr \ln [ 1-\exp (-\beta(\omegak  - \mu))] + \half \tr \ln [
 1-\exp (-\beta(\omegak  + \mu))]
 \label{F3def}
\end{eqnarray}
This has $\omegak  \pm \mu$ factors in temperature dependent terms
but not in the temperature independent terms, so is perhaps not
the most natural form to arrive at by direct calculation though
physically it is quite intuitive. The obvious way to obtain this
third form $F_3$ \tref{F3def} is by making algebraic manipulations
of the formal form \tref{F3mudef}. Later, inspired by the work of
\cite{MT}, I will find different answers in $\zeta$-function
regularisation for these three forms, \tref{F3mudef},
\tref{F3Ndef} and \tref{F3def}. Normal ordering gives a genuine
formal difference between the formal expressions $F_{3\mu}$
\tref{F3mudef} and $F_{3N}$ \tref{F3Ndef} but not between
$F_{3\mu}$ \tref{F3mudef} and $F_{3}$ \tref{F3def}. The difference
between $F_{3\mu}$ \tref{F3mudef} and $F_{3}$ \tref{F3def} is a
matter of the Multiplicative Anomaly
\begin{equation}\label{ashiftdef}
  a_{\rm shift} (A,\alpha) :=
  2\tr \{ A \} - \tr \{ A+\alpha \} - \tr \{ A-\alpha \}
\end{equation}
This expresses the failure of two basic algebraic identities
\tref{anomcshift} and \tref{anomcmult} discussed in appendix
\ref{appma}.   In the case of the three dimensional expressions
$F_{3}$ and $F_{3\mu}$, the difference is related to the
Multiplicative Anomaly \tref{ashiftdef} through
\beq
 F_{3} = F_{3\mu} + \half a_{\rm shift}(\omegak , \mu)
\eeq

Thus the unexpected differences between various $\zeta$-function
regularisation versions of the formally UV divergent canonical
calculations $F_{3}$ and $F_{3\mu}$, as found in the literature
and to be discussed below, are {\em not} caused by normal ordering
but failures of basic algebraic identities.  It is the difference
between $F_{3N}$ normal ordered UV finite expression and
$F_{3\mu}$ UV infinite expressions which is a matter of normal
ordering and the zero point energy in this model.  In this I
differ from both \cite{EFVZ} and \cite{MT98,MT} who emphasise the
role of normal ordering and canonical methods in their resolutions
for the problems.


\section{Free Bose gas results}

I will now turn from infinite formal expressions to finite
regularised ones.  Only with the latter can one compare the
results for physical quantities obtained using different
calculational schemes.  I will denote regularised quantities by
adding a further subscript: a $\zeta$ to indicated a
$\zeta$-regularised expression, or a $\epsilon$ to indicate that
dimensional regularisation was used, e.g.\ $F_{4K\zeta}$ is the
$\zeta$-function regularised form of $F_{4K}$.

Before looking at the results in detail, I will make a few
comments about the way I approached the four-dimensional finite
temperature calculations of \tref{F4Adef}, \tref{F4Kdef} and
\tref{F4Ldef} where there are energy variables to be summed over.
As noted above, I use a Euclidean approach to thermal field theory
where the energies are the discrete Matsubara energies but the
chemical potential is encoded directly in the Lagrangian,
propagators etc.\ rather than in the boundary conditions
\cite{LvW,TSEdal}.  There are standard methods for performing the
Euclidean energy sums, for instance using contour integration
methods.  In this case the integrands are logarithms or
non-integer powers of polynomials of energy so more care is needed
than with simple Green functions with their integrals of rationals
of polynomials. Nonetheless the discussion in the appendix
\ref{appftcalc}\tnote{??? (see also \cite{MT98})} shows that, as
with Green functions, one can separate any calculation into two
pieces
\begin{eqnarray}
  F(T ,\mu)
  &=& \int_\beta dk_4 \; g(k_4)
  \equiv \frac{1}{\beta} \sum_n g(k_4 = \frac{2 \pi n}{\beta})
  \label{gencalc0}
  \\
  F(T, \mu) &=& F_0(\mu)+ F_\beta(T ,\mu), \; \; \; \; \;
  F_0(\mu) = \int_{-\infty}^{+\infty} dk_4 \; g(k_4).
 \label{gencalc}
\end{eqnarray}
where in \tref{gencalc0} $g$ contains all the $\mu$ dependence. In
\tref{gencalc} $F_\beta$ contains {\em all} the explicit
temperature dependence in factors of $\exp \{ \beta (\omega \pm
\mu ) \}$\footnote{For Bose or Fermi statistics and $m > |\mu|$
such $F_\beta$ are always zero at zero temperature. However the
charge density is the physical parameter, not $\mu$, and this
makes $\mu$ an implicit function of temperature. See appendix
\ref{appftcalc} for further comments.}. The $F_\beta$ are also
guaranteed to be UV finite. The first piece, $F_0$, has no
explicit temperature dependence and is just the original
expression with the energy sum replaced by a Euclidean energy
integral from minus infinity to plus infinity. However, the zero
temperature piece $F_0$ will be chemical potential dependent. As
the problem of Multiplicative Anomalies is all about the behaviour
of UV divergences, the focus will be on the temperature
independent but possibly $\mu$ dependent UV divergent parts
$F_0(\mu)$. The major question is how to implement the
regularisation of the UV divergences in these temperature
independent, chemical potential dependent terms.

Various terms crop up again and again so I will define some
useful functions.\tnote{To relate to my notes I use $f_0 =
-W/(\beta V)$, $Y = -X/(\beta V)$, $f_\beta = -S/(\beta
V)$.}\tnote{!!! Had wrong sign in Y, is it now correct ???}
\begin{eqnarray}
 f_{-1}(m) := f_{-1} &=& -\frac{m^4}{32 \pi^2} ,
 \label{fm1def}
 \\
 f_{0}(m,M) :=  f_{0} &=&
  f_{-1}(m) \left[ \ln \left(\frac{M^2}{m^2}\right)
  +\psi(3) - \psi(1) \right]
  = \frac{m^4}{32 \pi^2} \left[ \ln \left(\frac{m^2}{M^2}\right) -
  \frac{3}{2} \right] ,
  \label{f0def}
  \\
 Y(m,\mu) := Y &=&  \frac{1}{16 \pi^2} \mu^2
                   \left( m^2 - \frac{\mu^2}{3} \right) ,
 \label{Ydef}
 \\
 X(m,\mu) := X &=&  \frac{Y(m,\mu)}{2 f_{-1}(m)}
 =  - \frac{\mu^2}{m^2}
    \left( 1- \frac{1}{3} \frac{\mu^2}{m^2} \right),
 \label{Xdef}
 \\
 f_\beta(m,\mu,\beta) :=  f_\beta &=&
 \frac{1}{\beta } \sum_\pm \intdtk
 \ln \left( 1- \exp \{ -\beta (\omegak  \pm \mu) \} \right) ,
 \label{fbetadef}
\end{eqnarray}


\subsection{4D $\zeta$-function regularisation results.}

The details about how to implement $\zeta$-function regularisation
in this simple case are given in appendix \ref{appzfr}.  I have
chosen to write these $\zeta$-function regularised expressions so
as to make clear the close relationship to the dimensional
regularisation results of later subsections. This will simplify
the comparisons I make in section \ref{discussion} but it means
that my $\zeta$-function regularisation expressions are not always
exactly the same as those in the literature, but the relationship
is trivial (see appendix \ref{appzfr}). My $\zeta$-function
regularisation versions of the formal expressions \tref{F4Adef},
\tref{F4Kdef} and \tref{F4Ldef} are
\bea
 F_{4A\zeta} &:= & \half \intdfk \left(-\frac{2}{s}\right)
 \left[ \frac{(k_4^2+\omegak^2+ \mu^2)^2 - 4 \omegak^2 \mu^2}{M^4}
 \right]^{-s/2}
 \label{F4Azdef}
 \\
 F_{4K\zeta} &:= & \half \sum_{\pm} \intdfk
 \left(-\frac{1}{s}\right)
 \left[ \frac{(k_4 \pm i\mu)^2+\omegak^2}{M^2} \right]^{-s}
 \label{F4Kzdef}
 \\
 F_{4L\zeta} &:= & \half \sum_{\pm} \intdfk
 \left(-\frac{1}{s}\right)
 \left[ \frac{k_4^2 + (\omegak \pm \mu)^2}{M^2} \right]^{-s}
 \label{F4Lzdef}
\eea
where I have chosen to ensure that in all expressions the
regularisation scale always appears as $M^{2s}$ and the first
term in a small $s$ expansion is always $O(1/s)$ to maintain a
close analogy with dimensional regularisation's $M^{2\epsilon}$
and $O(1/\epsilon)$. Using standard tricks it is straightforward
to do these integrals and I find for \tref{F4Kzdef} and
\tref{F4Lzdef}
\begin{eqnarray}
 F_{4K\zeta} &=& f_{-1}(m)\frac{1}{s} + f_0(m,M) + f_\beta(m,\mu,\beta)
 \label{F4Kzres}
 \\
 F_{4L\zeta} &=& f_{-1}(m)\frac{1}{s} + f_0(m,M)
 + 2 Y(m,\mu)
 + f_\beta(m,\mu,\beta)
 \label{F4Lzres}
\end{eqnarray}
Note that both of these are calculated directly from the
expressions quadratic in energy-momentum used to define them in
\tref{F4Kdef} and \tref{F4Ldef}, that is no use of Multiplicative
Anomalies was made in their calculation. The surprise is that
$F_{4L\zeta}$ is different from $F_{4K\zeta}$ by just the sort of
$\beta$ independent, $\mu$ dependent term I will be talking about
in the context of Multiplicative Anomalies. The expressions are
essentially the same\footnote{The only differences are the sign
in front of the $f_0$ term in \tref{F4Kzres} (equation (44) of
\cite{EFVZ}) and a factor of two in \tref{F4Lzres} (equation (38)
of \cite{EFVZ}).} as given in \cite{EFVZ}.

Now one can consider the expression $F_{4A\zeta}$ based on quartic
energy-momentum terms.  The calculation has been done in three
ways. First, one can work directly with the quartic terms and
\cite{EFVZ} obtain\tnote{Checked in MAPLE {\tt free Bose gas?.mws}
series.}
\begin{eqnarray}
 F_{4A\zeta} &=& f_{-1}(m)\frac{1}{s} + f_0(m,M)
 + Y(m,\mu)
 + f_\beta(m,\mu,\beta)
 \label{F4Azres}
\end{eqnarray}

In the second and third approaches to $F_{4A\zeta}$,  I start from
the definition of $F_{4A\zeta}$ in terms of $\Tr \ln$ over
quadratic operators and an additional anomaly term, as given in
\tref{F4AdefK} and \tref{F4AdefL}.   Results for the $\Tr \ln$
over quadratic operators were given above in \tref{F4Kzres} and
\tref{F4Lzres}. However, one sees from the definition of the
Multiplicative Anomaly \tref{adef} that to get the result for
$F_{4A\zeta}$ in these cases I must calculate the relevant
Multiplicative Anomalies, as indicated in \tref{F4AdefK} and
\tref{F4AdefL}. These types of Multiplicative Anomaly can be
calculated directly from a formula based on the Wodzicki residue
\cite{Wo} of the theory of elliptic pseudo-differential
operators.\tnote{It can also be deduced by reversing the flow of
logic in this section, but it was performed as stated here.}
Details are given in appendix \ref{appwod}. In any case, I find
that\tnote{Checked in MAPLE {\tt free Bose gas?.mws} series.}
\begin{equation}\label{anomres}
  a(K_+,K_-) = - a(L_+,L_-) = -2\beta V Y(m,\mu)
\end{equation}
which agrees with earlier results of Elizalde et al.\
\cite{EFVZ}.\tnote{\cite{EFVZ} (92), (96). Remember there is a
factor of $-1/2$ in relating the Multiplicative Anomalies in the
trace of the logarithm of the operators in \tref{F4AdefK} and
\tref{F4AdefL} to the free energy densities. This is where the
analysis of \cite{MT} version one goes wrong.  Also there is
missed divergence in another expression in \cite{MT}.} It is then
clear that all three methods of calculating $F_{4A\zeta}$ give the
same answer and therefore I disagree with the suggestion of an
inconsistency made in \cite{MT}.

The fact that all three approaches give the same answer is
fundamental. It is a matter of mathematical definition that all
three methods --- direct, via $K_\pm$ and its Multiplicative
Anomaly, and finally via $L_\pm$ and its Multiplicative Anomaly
--- are calculating the same object.
However, if we had forgotten the Multiplicative Anomaly then I
would have an inconsistency in our results for $F_{4A\zeta}$. In
this sense the existence of the Multiplicative Anomaly is not a
problem but is {\em absolutely essential} for mathematical
consistency.\footnote{Of course, if you accept that Anomalies
exist then this shows how to calculate them without knowing the
Wodzicki formula.  Calculate $F_{4A\zeta}$, and $F_{4K\zeta}$
directly as done in \cite{EFVZ}, and then one can deduce the
result for $a(K_+,K_-)$.}

Note that while there is complete consistency in the
$F_{4A\zeta}$ results, the $Y$ factor comes up in different ways
in different approaches.  When using $F_{4K\zeta}$ \tref{F4AdefK}
approach the $Y$ comes solely from the Multiplicative Anomaly
$a(K_+,K_-)$.  When starting from the $L$ based form $F_{4L\zeta}$
of \tref{F4AdefL}, there is a $Y$ term intrinsic to $F_{4L\zeta}$
and another one from the Multiplicative Anomaly $a(L_+,L_-)$
which contribute to the final $Y$ factor in $F_{4A\zeta}$
\tref{F4Azres}.

Of course, while mathematical consistency of the $\zeta$-function
regularisation may require non-zero Multiplicative Anomalies, this
appears to cause major problems to the physical interpretation.  I
will return to this later when all relevant results have been
acquired.

\subsection{3d $\zeta$-function regularisation results.}

The details of the finite temperature aspects and $\zeta$-function
regularisation are as for the four-dimensional calculations above
(also see appendices \ref{appftcalc} and \ref{appzfr}). Thus I
define the $\zeta$-function regularisation forms as
follows\tnote{From notes dated 29-7-00.}
\bea
 F_{3\zeta} &:=& \intdtk \; \omega^{1-2s} M^{2s}
 \label{F3zdef}
 \\
 F_{3mu\zeta} &:=& \half \sum_\pm \intdtk \; (\omega \pm \mu)^{1-2s} M^{2s}
 \label{F3muzdef}
\eea
and $F_{3N}$ is UV finite so needs no regularisation.  I choose
$s$ so that the renormalisation scale appears to the same power as
in the four-dimensional calculations.  Direct calculation then
gives
\bea
 F_{3N} &=& f_\beta
 \label{F3Nzres}
 \\
 F_{3\zeta} &=&
 f_{-1} \frac{1}{s}
   + f_{-1} \left(  \ln \left(\frac{M^2}{m^2}\right) +\psi(3) -\psi(-1/2)
   \right)
   + f_\beta
 \label{F3zres}
 \\
 &=& f_{-1} \frac{1}{s}
    + f_{0} + f_{-1} \left( \psi(1) -\psi(-1/2) \right) + f_\beta
 \\
 F_{3\mu\zeta} &=& F_{3\zeta} + 2Y(m,\mu)
 \label{F3muzres}
\eea
Thus I see that in $\zeta$-function regularisation
\begin{equation}
 a_{{\rm shift},\zeta} (\omega, \mu)
 = 2\tr(\omega) - \tr(\omega-\mu)- \tr(\omega+\mu)
 = 2 \left( F_{3\zeta} - F_{3\mu\zeta} \right)
 = -4\beta V Y(m,\mu).
\end{equation}
The Wodzicki residue formula \tref{wresanom} is only useful for
$\Tr \ln$ expressions.  Attempts to use it here forces one to try
exponentials of well behaved operators in the formula
\tref{wresanom} but these are not suitable. However, I am able to
get the form of this Multiplicative Anomaly directly, if not the
overall factor, by using a conjectured generalisation of the
Wodzicki residue formula \tref{anomconj}.

\subsection{4D dimensional regularisation results.}\label{4ddr}

It is useful to compare the $\zeta$-function regularisation
results against results of a method more often used in particle
physics, so I will use dimensional reduction. When using
dimensional regularisation at non-zero temperature it is important
to note that the regularisation takes place solely in the spatial
integration, as discussed in the appendix \ref{appdr}, i.e.\
\begin{equation}
 \int_\beta \dfek
 \equiv (2 \pi)^{2\epsilon-4} \frac{1}{\beta} \sum_{n} \int d^{3-2\epsilon}
 \label{drbetadef}
 \veck
\end{equation}
Let us start with the quartic expression \tref{F4Adef}. The
appropriate dimensionally regularised form at $T>0$ is
\bea
 F_{4A\epsilon}
 &:=&
 \int_\beta \dfek
 \;
 M^{2\epsilon} \ln \left(\frac{K_+K_-}{M^4}\right)
 \label{F4Aedef}
\eea
The key point here is that this expression is finite while
$\epsilon \neq 0$ so I can manipulate the integrand and swap the
order of summing integrands with integration, for instance
\bea
 F_{4K\epsilon} &:=&
 \int_\beta \dfek \; M^{2\epsilon}  \ln \left(\frac{K_+}{M^2}\right)
 +
 \int_\beta \dfek \; M^{2\epsilon}
 \ln \left(\frac{K_-}{M^2}\right)
 = F_{4A\epsilon}
  \label{F4KAe}
\eea
Thus it is immediately obvious that there are no Multiplicative
Anomalies in dimensional regularisation as noted in
\cite{TSEanom}. Denoting all calculations in dimensional
regularisation based on any of these four-dimensional forms as
$F_{4\epsilon}$ I find
\bea
F_{4\epsilon}     &=&
      f_{-1}(m) \frac{1}{\epsilon}
    + f_{-1}(m) \left( \ln\left(\frac{M^2}{m^2}\right)
                      + \ln (4 \pi ) + \psi(3) \right) +
    f_\beta(m,\mu,\beta)
    \label{F4eres}
\tpre{    \\
    &=&
      f_{-1}(m) \frac{1}{\epsilon}
    + f_{-1}(m) \left( \ln\left(\frac{4\pi e^\gamma M^2}{m^2}\right)
                      + \psi(3) -\psi(1) \right) +
    f_\beta(m,\mu,\beta)
    \label{F4eres2}
    }
    \\
    &=&
    f_{-1}(m) \frac{1}{\epsilon}
    + f_0(m,(4\pi e^\gamma)^{1/2}M)
    + f_\beta(m,\mu,\beta)
    \label{F4eres3}
\eea

Before moving on, note that is possible to produce failures of
algebraic identities in dimensional regularisation if one fails
to implement dimensional reduction correctly. For instance this
can be done by\tnote{I think this bit in this tnote is wrong:
altering the definition of a dimensionally regularised integral by
changing either by multiplying the c(d) by a function which is one
at $d=4$ and analytic near $d=4$, ...} regulating the energy sums
rather than the three-momentum integrals, see appendix \ref{appdr}
for further comments.

\subsection{3d dimensional regularisation results.}

Just as I noted for the four-dimensional calculations using
dimensional regularisation, if one follows the standard
application of the dimensional reduction scheme to one of the
three dimensional forms \tref{drdef} one quickly sees that there
is no problem with simple algebraic identities such as
\tref{anomcshift} and hence the Multiplicative Anomaly $a_{\rm
shift}$ of \tref{ashiftdef} is zero in dimensional reduction. Thus
there is no difference between the two dimensionally regularised
expressions based on the symmetric ordering, $F_{3\mu}$ of
\tref{F3mudef} and $F_{3}$ and \tref{F3def}. Likewise, the third
expression, $F_{3N\epsilon}$ is just the UV finite $T>0$ term
present in the all the other expressions encountered in
dimensional reduction or $\zeta$-function regularisation
calculations. The dimensionally regulated form of both
\tref{F3def} and \tref{F3mudef} is then
\bea
 F_{3\mu \epsilon} =
 F_{3\epsilon} &:=& \int \dslash^{3-2\epsilon} \veck \;
 M^{2\epsilon} \; \omega(k) ,
 \; \; \;
 F_{3N\epsilon} = f_\beta
 \label{F3edef} \label{F3muedef}
\eea
From this I find
\bea
 F_{3\epsilon}(m,M_\epsilon,\mu,T,\epsilon)
 &=&
  f_{-1}(m) \frac{1}{\epsilon}
 + f_{-1} (m) \left(   \ln \left(\frac{M_\epsilon^2}{m^2}\right) + \ln(4\pi^2)
                 + \psi(3) \right)
 + f_\beta
 \label{F3eres}
 \label{F3mueres}
\eea
We see that the three- and four-dimensional results (except for
the UV finite $F_{3N\epsilon}$) are identical with the {\em same}
choice of $\epsilon$ and renormalisation scale $M_\epsilon$. From
this I find
\bea
 F_{3\epsilon}(m,M_\epsilon,\mu,T,\epsilon)
 &=&
 F_{4\epsilon}(m,M_\epsilon,\mu,T,\epsilon)
\eea
This is not entirely trivial but comes as a result of the careful
implementation of the dimensionally regularised scheme, e.g.\
before integration free energies are always proportional to
$M^{2\epsilon}$. This ensures identities such as \tref{drprod} are
always satisfied which helps link three- and four-dimensional
results. One can also make this connection directly. First,
remember that $F_{4A\epsilon}$ is regulated only in the spatial
integration, as shown in \tref{F4Aedef}.\tnote{???}  Doing the
Euclidean energy sum is thus straightforward using the usual
contour integration methods as all the regularisation is in the
spatial momentum. This shows that there are no $T=0$ $\mu$
dependent terms, just a $T=0$ UV divergent piece and the UV finite
$T>0$ contribution $f_\beta$.\footnote{The $T=0$ part of
$F_{4K\zeta}$ can be rewritten using the identity \tref{drprod} as
the usual four-dimensional zero point energy result for a complex
scalar field in dimensional regularisation.}  More relevant is
that the $T=0$ piece of $F_{4\epsilon}$ is then a
three-dimensional integral and is then clearly of the form
\tref{F3edef},\tnote{???} the three-dimensional expression
$F_{3\epsilon}$.

\section{Comparison of results}\label{discussion}

I have focused on six different expressions using two different
regularisation schemes all purportedly for the same quantity - the
free energy of a free charged Bose gas. As always in QFT,
comparison of results is complicated by the need to make sure that
in two expressions for the same quantity that the physical
parameters, upon which the answer depends, are defined in the same
manner in all cases.  Since it is a free field theory, there are
no obvious corrections to the bare mass which is the physical
mass, despite the quantum and statistical fluctuations encoded in
the calculations.\tnote{What about the normalisation of the
propagator? It does not factorize in $\zeta$-function
regularisation so does this cause problems?} However one also has
the quartic UV divergences in the free energy density or
equivalently in the zero point energy density.\tnote{In the case
of the free energy of the free Bose gas, it has a quartic
divergence and an infinite subtraction is required, even though
there are no interactions and no renormalisation of the mass seems
to be required.Normalisation of propagator when $Z$ won't
factorize?}  Thus to discuss the physics in these results one must
not only discuss the {\em regularisation} of the UV divergences,
but also the removal of these divergences as part of the {\em
renormalisation} procedure\tprenote{The latter process is needed
when extracting the physics in QFT whether or not there are UV
divergences.}.  I will postpone the discussion of the renormalised
results and the physics to the following section and in this
section I will compare the full regulated results.

The regulated but unrenormalised $\zeta$-function regularisation
and dimensional reduction results can be compared directly by
setting $\epsilon = s$ as by inspection one see that all the
singular terms are then identical. This is not a miracle but comes
from the close relationship between dimensional reduction and
$\zeta$-function regularisation and from the way I have chosen
various optional factors in such a way as to ensure this
behaviour, see the appendices for further details\footnote{These
extra factors, functions $g$ in \tref{cddef} or \tref{zfrlnop} in
the appendices, can easily be included provided one modifies the
relationship to $s=\epsilon / g(\epsilon)$ or similar. Fixing
$g=1$ still leaves the freedom to manipulate the renormalisation
scale $M$.  Thus I choose to work with $g=1$ merely to remove
irrelevant complications.}. Also note that the expressions for the
Multiplicative Anomalies are themselves na\"{\i}vely finite and in
principle require no renormalisation to obtain finite answers,
though one needs to implement regularisation carefully to ensure
this is seen in the final answers.\tnote{Had one compared with
integrals regulated by cutting off the momentum integrals at
$\Lambda$, one would have to compare $m^4/\epsilon$ against a
combination of $\Lambda^4$ and $m^2\Lambda^2$ terms, giving a
complicated and mass dependent relationship.}

However, regularisation alone introduces a new scale, $M$ the
renormalisation scale.  This is not a trivial object.\tnote{In an
ideal world one would always use physical renormalisation
conditions (e.g.\ as defined in Cheng and Li) to ensure all
parameters were physical.  $M$ would then be a physical scale,
e.g.\ one would define the mass at zero temperature at a scale
equal to the physical mass, a legitimate if self-consistent
definition. However even in this case $M$ will depend on the
regularisation scheme employed.  In practice though, the situation
is further complicated through the use of unphysical but easily
implementable renomalisation schemes.} A good practical example of
the importance of $M$ is the way that in particle physics, a great
deal of effort goes into setting the scale for lattice and
$\overline{{\rm MS}}$ dimensional regularisation results for the
same quantities \cite{PDG}.\tnote{For instance in ${\rm MS}$
renomalisation scheme of dimensional reduction, or in traditional
implementations of the $\zeta$-function regularisation scheme, one
just drops the leading term in $\epsilon$ or $s$ which represents
the UV divergences.  Thus counter terms are just being defined
rather than being expressed in terms of some physical observable
value.}\tnote{In principle, the value of $M$ should be justified
by comparing our na\"{\i}vely UV divergent but regulated
calculation to some known finite physical value. This leaves open
the possibility that when I subtract the UV divergences in
different calculations, I leave different finite pieces behind in
our physical results. This is quite normal and is normally
equivalent to different definitions of the physical quantity, and
this in turn is related to the fact that in different
calculational methods for the same quantity the renormalisation
scales {\em must} be different.}

The conclusion is that in order to be able to compare the
different results for the free energy obtained above I must not
assume that the renormalisation scales $M$ appearing in different
calculations are the same.

\subsection{Discussion of dimensional regularised results.}

As I will consider these to represent the `standard' results, I
will discuss these first. Several aspects are well known but, in
view of the complications associated with the $\zeta$-function
regularisation results, I will repeat them and emphasize several
aspects for later comparison with the $\zeta$-function
regularisation results.

The canonical expression for the normal ordered expression
$F_{3N}$ leads to the non-zero temperature term $f_\beta$.  This
is UV finite and hence independent of regularisation. $f_\beta$ is
also zero at zero temperature.\tnote{For bosonic or fermionic
fields this is always true for all $|\mu |<m$.}  This accords with
the usual idea that normal ordering removes the zero point energy
term associated with expectation values in the pure vacuum state.
Thus $f_\beta$ appears to be the $T>0$ UV finite correction to the
zero point energy, and indeed we will find this borne out below in
other non-normal ordered calculations. The finite temperature
literature usually focuses on this term alone and leaves the
temperature independent UV divergent zero point energy terms to
one side e.g.\ \cite{Ka}.

The other five expressions for the free energy density, which are
all UV divergent, lead to identical results in dimensional
regularisation. Specifically the four-dimensional results
$F_{4A\epsilon} \tref{F4Aedef}, F_{4K\epsilon}$ \tref{F4KAe} and
$F_{4L\epsilon}$ lead to \tref{F4eres}.  The two three-dimensional
expressions $F_{3\epsilon}$ and $F_{3\mu\epsilon}$ of
\tref{F3edef} and \tref{F3mudef} respectively lead to the single
result \tref{F3eres}, which is identical to the four-dimensional
dimensional reduction result \tref{F4eres} if the same
renormalisation scale $M$ is used\footnote{The fact that the
three- and four-dimensional dimensional regularisation results are
equal with the same renormalisation scale follows from identities
in dimensional regularisation such as \tref{drprod}.}. Since
equations such as \tref{F4AdefK} show that any differences in
results are related to non-zero anomalies, my conclusion is that
there are there are no Multiplicative Anomalies in dimensional
regularisation of the free Bose gas model.  This agrees with the
suggestion made in the introduction, based on the criteria set out
in \cite{TSEanom}, namely that conventional dimensional reduction
is always Multiplicative Anomaly free.

The result for these UV divergent expressions comes in two parts.
First, the difference between $T>0$ and $T=0$ values (given that
$|\mu |<m$) is always the same $f_\beta$ term as in
$F_{3\epsilon}$. The remaining part, $r^{-1}f_{-1} + f_0$ (where
$r$ is either $s$ or $\epsilon$) is independent of temperature,
and contains both finite contributions and UV divergences.  An
important point is that in dimensional regularisation this $T=0$
part is also independent of chemical potential. Since the normal
ordered form $F_{3N}$ should have removed the zero point energy,
$(F_{3}-F_{3\mu})$ is the zero point energy in dimensional
reduction. The zero point energy in dimensional reduction is made
up of the temperature and chemical independent terms (infinite and
finite) as one would expect from its definition in terms of a
vacuum energy expectation value\footnote{One may easily work at
$T=0$ and $\mu \neq 0$ to confirm this directly, as appendix
\ref{appftcalc} shows.}. Thus normal ordering merely removes the
UV divergent zero point energy, it is not the source of any
Multiplicative Anomaly. Indeed there are no Multiplicative
Anomalies in dimensional reduction \cite{TSEanom}, as discussed
earlier, while normal ordering has the expected effect in
dimensional reduction, clearly showing there is no link between
Multiplicative Anomalies and normal ordering. If any final proof
of this is needed, recall that the appropriate normal ordering at
$T>0$, $\mu \neq 0$ is not the usual one considered at zero
temperature \cite{ES}.\tnote{ It also follows then that while
there is no obvious operator ordering encoded in the path integral
derived four-dimensional expressions, the path integral tends to
chose a symmetric ordering such as was used for the
three-dimensional forms, \tref{F3def} etc., given that all five UV
divergent results are identical.}

One last comment upon the dimensional regularisation results is to
note that all of the UV divergent forms in dimensional
regularisation are equal when the same renormalisation scale,
$M_\epsilon$, is used. This is a direct consequence of the
consistency demanded for dimensional regularisation integrals
\cite{Co} and in particular comes from the result \tref{drprod}
which can be used to link the three and four-dimensional results
directly with the same scale $M_\epsilon$.

\subsection{Discussion of $\zeta$-function regularisation results}

The $\zeta$-function regularisation produces the same result as
dimensional reduction for the UV finite normal ordered expression
$F_{3N\zeta}$. After this, things become more complicated.

Let us start with the four dimensional form $F_{4K}$.  The
$\zeta$-function regularisation result for $F_{4K}$ has no terms
which depend on chemical potential other than the UV finite
$f_\beta$, e.g.\ no $Y$ factors. This is the same behaviour as all
the dimensional regularisation results.  One can see this result
quickly when using the Schwinger trick
\begin{equation}\label{schwinger}
  \frac{1}{a^s} = \frac{1}{\Gamma(s)} \int_0^\infty dt \; t^{s-1}
  e^{-at}
\end{equation}
on \tref{F4Kzdef}.  There the $T=0$ term has an energy integral
which becomes a Gaussian with a peak at $\mu$ so a simple shift
removes all $\mu$ dependence from the $T=0$ terms. By inspecting
the results for $F_{4K\zeta}$ it is therefore straight forward to
make a link between the dimensional regularisation results and
this $\zeta$-function regularisation result.  From \tref{F4eres}
equality is obtained by choosing slightly different renomalisation
scales for this $\zeta$-function regularisation calculation and
the dimensional reduction calculations\tnote{Of course the need
for different renormalisation scales in different regularisation
and renormalisation schemes to obtain the same physics is
completely normal though for interacting theories it is both
highly non-trivial and of great practical importance e.g. when
comparing lattice and $\overline{MS}$ dimensional regularisation
calculations for particle physics experiments \cite{PDG}.}
\bea
F_{\epsilon} (m,M_{\epsilon},\mu,T,\epsilon=s) &=&
F_{4K\zeta}(m,M_{4K\zeta},\mu,T,s=\epsilon) ,
\\
(4 \pi e^\gamma)^{1/2} M_\epsilon &=& M_{4K\zeta}
\eea

The problem is with the other two four-dimensional
$\zeta$-function regularisation results, $F_{4A\zeta}$ of
\tref{F4Azres} and $F_{4L\zeta}$ of \tref{F4Lzres}.  Both differ
from $F_{4K\zeta}$ by factors of $Y$, which is both mass and
chemical potential dependent and which can not be absorbed as
constant shifts to the regularisation scale. In fact I see that
to obtain equality I must set
\bea
 F_{4A\zeta} (m,M_{4A\zeta},\mu,T,\epsilon=s) &=&
 F_{4K\zeta}(m,M_{4K\zeta},\mu,T,s=\epsilon) ,
 \\
 e^{X} M_{4A\zeta} &=& M_{4K\zeta}
 \\ F_{4L\zeta}(m,M_{4L\zeta},\mu,T,\epsilon=s) &=&
 F_{4K\zeta}(m,M_{4K\zeta},\mu,T,s=\epsilon) ,
 \\
 e^{2X} M_{4L\zeta} &=& M_{4K\zeta}
\eea
where $X(m,\mu)$ is defined in \tref{Xdef}.  Thus as $0 \leq |\mu|
\leq m$ I have\tnote{$0 \leq X \leq - 2/3$,} $1 \geq e^X \gtrsim
0.513$ and $1 \geq e^{2X} \gtrsim 0.264$, sizable shifts in the
renormalisation scales if I want to demand equality of free
energies in these different $\zeta$-function regularisation
schemes.\tnote{Finally remember that using Multiplicative
Anomalies one can relate all these four-dimensional
$\zeta$-function regularisation calculations through equations
\tref{F4AdefK} and \tref{F4AdefL}, and by implication one then has
complete mathematical consistency.}

Having dealt with the four-dimensional forms, let me now turn to
the three dimensional results using $\zeta$-function
regularisation \tref{F3zres} and \tref{F3muzres}. Comparing with
previous $\zeta$-function regularised results for the
four-dimensional calculations, I see that
\bea
 F_{3\zeta}   (m,M_{3\zeta},\mu,T,s) &=& F_{4K\zeta}(m,M_{4\zeta},\mu,T,s) ,
 \label{F34z1}
 \\
 F_{3\mu\zeta}(m,M_{3\zeta},\mu,T,s) &=& F_{4L\zeta}(m,M_{4\zeta},\mu,T,s)
 \label{F34z2}
 \eea
provided I shift the renormalisation scale for these
three-dimensional $\zeta$-function regularisation results as
compared with the four-dimensional scales as
\beq
\tpre{M_{3\zeta}^2 =  M_{4\zeta}^2 \exp \{ \psi(-1/2) -\psi(1) \}
 ,  \; \Rightarrow \;}
 M_{3\zeta}   = \frac{e}{2} M_{4\zeta} .
 \eeq
In fact one can derive these relationships directly from the
four-dimensional forms, and in doing so prove explicitly the
relationships \tref{F34z1} and \tref{F34z2} between the three- and
four-dimensional $\zeta$-function regularisation expressions. This
can be done by starting with the four-dimensional form, rewriting
the integrand using the Schwinger trick \tref{schwinger}, then
doing the energy integration, and finally inverting the Schwinger
trick \tref{schwinger}.

However one tackles the problem, the results \tref{F34z1} and
\tref{F34z2} show that the three-dimensional $\zeta$-function
regularisation forms, as used in \cite{MT98,MT}, hold no new
lessons as compared to the four-dimensional case, which were the
focus in \cite{EFVZ}.  As in dimensional reduction, normal
ordering gives a difference between normal ordered $F_{3N\zeta}$
and symmetric ordered three-dimensional $F_{3\zeta}$
$\zeta$-function regularisation expressions, specifically it
removes the UV divergences.  However the Multiplicative Anomaly
problem comes in comparing $F_{3\zeta}$ and $F_{3\mu\zeta}$, both
based on the same symmetric ordering of operators and so any
difference is {\em not} due to operator ordering.  Since the
three- and four-dimensional results can be linked by \tref{F34z1}
and \tref{F34z2}, the conclusion is that operator ordering can not
be responsible for the differences in UV divergent
four-dimensional expressions either. The slightly more complicated
relationship between the renormalisation scales in three- and
four-dimensional calculations is to be expected in general
regularisation schemes, and was only avoided in dimensional
reduction because of identities such as \tref{drprod}.

Still, in $\zeta$-function regularisation it appears that one can
obtain a single result for the UV divergent expressions only if we
allow a rescaling of the renormalisation scale that depends on
both mass and chemical potential.  By comparison in dimensional
reduction simple equality of all UV divergent results was
achieved.

Having shown how one can link all these different results, I will
now turn to the most important question, namely are there any
physical differences encoded by these results.

\section{The physics of the free Bose gas}

The results of the previous section are summarised in table
\ref{comptable}.
\begin{center}
\begin{table}[htb]
\begin{tabular}{c||c|c|c||c}
  Quantity & Coeff. of            & Coeff. of          & Coeff. of   & Relative\\
           & $r^{-1}f_{-1}+f_0$   & Y                  & $f_\beta$   & scale used\\
           & ($T=0$ $\mu$ ind.)   & ($T=0$ $\mu$ dep.) & ($T>0$)     & \\
\hline \hline
  Generic $F$              & a & b &  c & z \\
\hline
  $F_{3N\epsilon}$
  and $F_{3N\zeta}$        & 0 & 0 &  1 & -\\
  All other $F_{\epsilon}$ & 1 & 0 &  1 & $(4\pi e^\gamma)^{-1/2}$\\
  $F_{4K\zeta}$            & 1 & 0 &  1 & 1 \\
  $F_{4L\zeta}$            & 1 & 2 &  1 & $e^{-2X}$ \\
  $F_{4A\zeta}$            & 1 & 1 &  1 & $e^{-X}$ \\
  $F_{3\zeta}$             & 1 & 0 &  1 & $e/2$ \\
  $F_{3\mu\zeta}$          & 1 & 2 &  1 & $e^{1-2X}/2$
\end{tabular}
 \caption{Table of the different terms appearing in each expression for the free energy.
 The necessary rescalings of the renormalisation scale relative to the $F_{4K\zeta}$
 calculation scale are given.  The generic example is therefore
 $F= a(r^{-1}f_{-1}+f_0) + b Y + cf_\beta$ using a scale $M = z M_{4K\zeta}$.
 $r=s$ or $\epsilon$ as appropriate.}
 \label{comptable}
\end{table}
\end{center}
However, before any discussion of physics in these results can
take place, one must renormalise the free energy density, $F$.  I
will consider two distinct methods of removing its quartic
divergences, first a physical subtraction and then an unphysical
minimal subtraction.

\subsection{Physical Subtraction}

Consider a free energy difference, $\Delta F$, defined with
respect to a reference temperature $T_R$ and chemical potential
$\mu_R$, all other physical parameters held fixed
\begin{equation}\label{DFdef}
  \Delta F(m,M,T,\mu) = F(m,M,T,\mu) - F(m,M,T_R,\mu_R) .
\end{equation}
Such a subtraction will not alter the thermodynamics and it is
common to work with energy differences in many problems. One
quickly finds that there are three distinct results for $\Delta
F$. For all the dimensional regularisation calculations, and for
the $\zeta$-function regularisation calculations based on
$F_{4K\zeta}$, $F_{3N\zeta}$ and $F_{3\zeta}$ I have
\begin{equation}\label{DF1}
 \Delta F_{\epsilon} := \Delta F_{4K\zeta} = \Delta F_{3N\zeta} = \Delta F_{3\zeta}
  = f_\beta(m,T,\mu) - f_\beta(m,T_R,\mu_R)
\end{equation}
where $f_\beta$ of \tref{fbetadef} is the usual $T>0$ term
encountered in the free Bose gas model e.g.\ in \cite{Ka}.  The
$\zeta$-function regularisation result based on the quartic
operator, $F_{4A\zeta}$, picks up an extra $Y$ difference term
\begin{equation}\label{DF2}
 \Delta F_{4A\zeta}   = \Delta F_{\epsilon}  + Y(m,\mu) -
 Y(m,\mu_R),
\end{equation}
while the $L$ factorisation, $F_{4L\zeta}$, and one of the
three-dimensional results, $F_{3\mu\zeta}$, in $\zeta$-function
regularisation give
\begin{equation}\label{DF3}
 \Delta F_{4L\zeta}   = \Delta F_{3\mu\zeta} = \Delta F_{\epsilon}  + 2Y(m,\mu) - 2Y(m,\mu_R)
 .
\end{equation}
These are all UV finite, though in most interacting models a
single subtraction would not normally leave a finite result in
this way. They are also all independent of the renormalisation
scale $M$, provided we use the same value for $M$ at $T$ and $\mu$
as at the reference point $T_R,\mu_R$.  One can check\tnote{Do so}
that the physical quantities, various derivatives of $\Delta F$,
as a function of the observables, $T$ and the charge density $Q/V$
rather than a function of $T$ and $\mu$, do differ depending which
result from \tref{DF1}, \tref{DF2} and \tref{DF3} we take.  They
are though independent of the reference point $T_R, \mu_R$. At the
same time, all the different definitions of $F$ given in section
\tref{formaldef} seem to be equally good. Thus there appears to be
a serious problem in identifying the physics even in this simple
free Bose gas model. Before suggesting a resolution, let me now
consider a different renormalisation scheme.

\subsection{Minimal subtraction}

In minimal subtraction, one just drops the divergent terms. This
is a common scheme in dimensional reduction and it is also
performed implicitly in $\zeta$-function regularisation as
explained in appendix \ref{appzfr}.  Now I find four distinct
results, which I denote as $F^{\rm MS}$.  The UV finite normal
ordered results are unchanged by this renormalisation
\begin{eqnarray}
 F^{\rm MS}_{3N\epsilon} = F^{\rm MS}_{3N\zeta}
 &=& f_\beta(m,T,\mu) .
\end{eqnarray}
Otherwise, the results fall into three types depending on the
calculational scheme, just as they did for the physical
subtraction renormalisation used above.  Thus
\begin{eqnarray}
 F^{\rm MS}_{3\epsilon} = F^{\rm MS}_{3\mu\epsilon} = F^{\rm MS}_{4\epsilon} =
 F^{\rm MS}_{4K\zeta}
 &=& f_0(m,M) + f_\beta (m,T,\mu)
\\
 F^{\rm MS}_{4L\zeta}
 &=& f_0(m,M) + f_\beta (m,T,\mu) + Y(m,\mu)
\\
 F^{\rm MS}_{4A\zeta} = F^{\rm MS}_{3\mu\zeta}
 &=& f_0(m,M) + f_\beta(m,T,\mu) + 2Y(m,\mu)
\end{eqnarray}
All these renormalised free energies based on UV divergent
expressions are finite again but now all have a factor of $f_0$
\tref{f0def} unlike in the case of the physical subtraction
results.  Thus these all depend on the renormalisation scale $M$.
However the $f_0$ factors do not alter the thermodynamics,
assuming the same $M$ is used at all $T$ and $\mu$, so these
minimal subtraction calculations give the same physical results as
when the physical subtraction renormalisation is used.  The
physical results appear to be independent of the renormalisation
scheme yet depend on the details of the definition and
regularisation of the free energy.

\subsection{Resolution of the paradox}

The clue comes from the minimal subtraction results which retain
some explicit $M$ dependence even though this does not effect the
physical thermodynamics.  It is a reminder that when comparing
results at different $T$ and $\mu$ (or even different $m$), we are
naturally inclined to hold this renormalisation scale constant.
This is implicit in the physical subtraction calculations
\tref{DF1}, \tref{DF2} and \tref{DF3} where $M$ is assumed to be
the same at $T,\mu$ as at the reference point $T_R,\mu_R$.

But why should the renormalisation scale $M$ be held constant?
Well it first appears when we cutoff the UV modes at some scale
$\Lambda$ using a regulating function $R(k)$.  In the language of
this paper, $\Lambda \sim M/r$ where $r=\epsilon$ in dimensional
reduction and $r=s$ in $\zeta$-function regularisation.  Even
though the interesting physics will be happening at scales much
less than $\Lambda$ and will be encoded in $O(r^0)$ terms of a
small $r$ series, there will also be a memory of the UV modes in
the $O(r^0)$ terms and these will be $M$ dependent.

Now if we compare a regularised free energy expression for
different physical parameter values, e.g. calculate $\Delta F$, it
is absolutely essential that we have subtracted these UV modes in
{\em exactly} the same manner in both expressions. If not each
expression will have a slightly different remnant of the UV modes
and the difference in the way the UV sector was treated will
appear in the physical results and may be finite. It is though a
purely mathematical artifact, a sign that we did not deal with the
UV sector carefully enough and that we are not comparing like with
like.

For instance, the simplest way to cut off the UV modes is to use a
straight cutoff, inserting a regulating function $R_\lambda =
\theta(M/\lambda - |\veck|)$ ($\lambda \rightarrow 0$) into all
loop integrals\footnote{At non-zero temperature, the best approach
is to cutoff only the three-momentum, but such detail is not
relevant to the discussion here.}. This will give terms such as
\begin{equation}\label{Fldef}
  F_\lambda = \intdfk R_\lambda (k) \ln(k^2+m^2)  =
  c M^4 \frac{1}{\lambda^2}
  +\ldots
\end{equation}
where $c$ is some number.  The UV cutoff scale is at $\Lambda = M/
\lambda^{1/2}$ where $\lambda$ is the small parameter playing the
same role as $\epsilon$ in dimensional reduction and $s$ in
$\zeta$-function regularisation.   There is then no doubt that we
are removing the UV modes in the same manner for all calculations
and, provided that the physics was occurring at scales much less
than $\Lambda=M/\lambda^{1/2}$, no physics would be effected. If
we were then to compare expressions for the free energy at
different temperature or chemical potential in the free Bose gas
model, e.g.\ calculate $\Delta F$, the expected result of
\cite{Ka} would appear, namely it would depend only on $f_\beta$
factors and no $cM^4\lambda^{-2}$ type term would survive, no
extra $T$ or $\mu$ dependent terms.\tnote{I am mixing two ideas.
First $\Lambda=\Lambda(\mu)$ vs. different UV contributions in
different formal equiv. expressions.   The latter leads to former
but not necessarily vice versa.}

Suppose though that one chose the cutoff $\Lambda (\mu)$ to have
some slight dependence on the chemical potential, e.g.\ $\Lambda^4
(\mu) = M^4/\lambda^2 + \mu^4 $ (or indeed any other physical
parameter). If one compared the same regulated expression at two
slightly different $\mu$ values (e.g. looking at a $\mu$
derivative of the free energy needed to calculate the physical
charge density), terms depending on $\Lambda$ would not exactly
cancel. For instance making a physical subtraction we would find
$\Delta F_\Lambda \simeq c(\mu^4-\mu_R^4) + \ldots$ for the
simple example mentioned above. A finite remnant of the UV mode
contribution is left, it is chemical potential dependent and it
therefore alters the thermodynamics.

Thus in the case of a simple cutoff, the regularisation procedure
is obvious. The cutoff $\Lambda$ must be taken to be independent
of physical parameters.\tnote{To compare with the other
renormalisation schemes, write $\Lambda = M/ \lambda$ where
$\lambda$ is the small parameter playing the same role as
$\epsilon$ in dimensional reduction and $s$ in $\zeta$-function
regularisation.  The dimensionless small parameters are chosen
with great care to ensure the UV divergences cancel in expressions
which are naively finite, such as $\Delta F$, as appendices
\ref{appdr} and \ref{appzfr} discuss. The only freedom in the
choice of cutoff scale is in the dimensionful parameter $M$.  I
now don't think that $\Delta F_\Lambda \simeq
4c(\mu-\mu_R)\partial(\ln(M))/\partial \mu + \ldots$ would be the
extra term left coming from energy scales of order $\Lambda = M/
\lambda^{1/2}$ if the scale $\Lambda$ or equivalently $M$ was
chosen to be $\mu$ dependent.}

In the case of a simple cutoff, making the cutoff $\Lambda$, or
equivalently $M$, depend on physical parameters is clearly a bad
idea but it is easily noted and avoided. However, what if the {\em
form} of the cutoff function $R$ was $\mu$ dependent rather than
the cutoff scale itself? For instance
\begin{equation}\label{Rvardef}
  R_\lambda = 2 [ \exp \{ (|\veck| - M/\lambda)/w \} + 1 ]^{-1}
\end{equation}
is a theta function with its sharp jump at $M/\lambda$ smoothed
over a region of size $w$.  If the width $w$ varies slightly with
$\mu$ then this will generate spurious $\mu$ dependent terms from
high energy $k \sim M/\lambda$ modes when comparing free energy
calculations at different chemical potentials.

Of course one could try to combine these two ideas.  Thus one can
compensate for a cutoff function $R$ which depends on $\mu$ by
choosing the scale $M$ to be $\mu$ dependent too, chosen so that
these spurious $\mu$ dependent terms from $M/\lambda$ modes in any
physical result were removed. However it is extremely difficult to
do this and $M$ would have to be a complicated function of $\mu$.
In fact the only easy way to distinguish physical terms from
spurious ones coming from $M/\lambda$ modes is to compare against
a calculation done with a regulating function $R$ and
renormalisation scale $M$, both of which chosen constant.  In this
case though it is clear that the one may as well just work with
the fixed regulator prescription in the first place.

Armed with these simple examples in terms of the cutoff $\Lambda
=M/\lambda^{1/2}$, let me now return to the main calculations of
this paper. The discussion in the previous section
\ref{discussion} showed how to use the renormalisation scale $M$
to relate all the different results but in several cases $M$ had
to depend on the physical parameter $\mu$. This suggests that
what is happening in the different calculations is that the UV
modes around the scale $M/r$ are being removed in different ways
for different physical parameter values.  Drawing on the comments
in the appendix \ref{appdr}, it is clear that dimensional
reduction is not the offender. Roughly speaking, dimensional
reduction regulates by inserting a function $R_\epsilon =
(\veck^2/M^2)^{-\epsilon}$, which is completely independent of any
physical parameters.

However, $\zeta$-function regularisation regulates by altering the
integrands.  For instance for the $L$ factorization, the regulator
is, crudely speaking, $R_\zeta = [(k_4^2 + (\omega\pm
\mu)^2)/M^2]^{-s}$ in \tref{F4Lzdef}.  This is clearly changing
the way the UV modes are cutoff in a manner which depends on
$\mu$ and $m$. It is therefore no wonder that strange $\mu$ and
$m$ dependent terms, $Y$, appear even in the physical results.
Equally, it should be no surprise that by choosing $M$ to be a
suitable, but complicated, function of $\mu$ and $m$, such terms
can be removed.

The lesson from the $\Lambda$ cutoff examples is that one should
not trust dependence on physical parameters in calculations using
regulators which involve those same physical parameters. Thus in
$\zeta$-function regularisation one should not trust the $\mu$ or
$m$ dependence of the results for the free energy of a free Bose
gas. Also, the only way to identify the true physics is to use a
regulation scheme which is not sensitive to the physical
parameters.  Therefore one can not ascribe some fundamental
meaning to one or other of the $\zeta$-function regularised forms
results ({\em pace} \cite{MT98,MT}). In the case of the free Bose
gas one may as well just work with a simple constant UV cutoff or
dimensional reduction as $\zeta$-function regularisation merely
adds unnecessary complications.

\section{Conclusions}

My first conclusion is to confirm the assertion of \cite{EFVZ}
that if and only if one includes Multiplicative Anomalies in
$\zeta$-function regularisation calculations does one get
mathematical consistency in the free Bose gas model.  That is if
one considers a well defined mathematical object, such as the
$\zeta$-function {\em regulated} expression for $F_{4A}$
\tref{F4Azdef}, it does not matter then how you calculate it,
either directly or via other quadratic expressions with the
appropriate Multiplicative Anomalies, the same answer is always
obtained. This is in contrast to suggestions elsewhere \cite{MT}
that different calculational approaches to one well defined
mathematical object in $\zeta$-function regularisation might give
different answers because of some subtle physical effect.  This
point has been confirmed in several other models, see
\cite{EVZ,TSEanom,FSZ,Su} and appendix \ref{appma} for more
details.

The second conclusion of this work is that the $\zeta$-function
regularisation calculations of the free Bose gas can not be
trusted. Perhaps after a careful study of the regulation of the UV
modes, one might be able to be sure that some $\zeta$-function
regularisation results are free of spurious finite UV mode
contributions. However it is far easier to use an physics
independent UV regulation scheme, such as dimensional reduction,
to find the correct physics in this case. Thus the thermodynamics
of the free Bose gas is described by $f_\beta(m,T,\mu)$ of
\tref{fbetadef} alone, which is the standard result of the
literature e.g.\ \cite{Ka}. The additional terms, such as
$Y(m,\mu)$ of \tref{Ydef}, found in some results, such as
$F_{4A\zeta}$, are genuine enough mathematically but they do not
have a simple physical interpretation, they are merely regulation
artifacts contrary to suggestions made  in \cite{EFVZ,Fi}.

This second conclusion means that one should not worry about the
physics obtained when such artifacts are included in the free
energy, as was done in \cite{EFVZ}.  All the different results,
$F_{4K\zeta},F_{4A\zeta}$ etc., all encode the physics of this
model exactly but they do so in an extremely convoluted manner.
Thus I also disagree with \cite{MT98,MT} that only some of these
expressions are physical, e.g.\ $F_{4K\zeta}$, and the others,
such as $F_{4A\zeta}$, are unphysical. The dimensional
regularisation calculations confirm that all the formal starting
points are equally good.  It is merely that some regularisation
schemes are less convenient than others.\tnote{In those schemes
one can only easily extract the physics by comparing with results
obtained in a more convenient regularisation scheme.} Thus only by
comparing $\zeta$-function regularised calculations against
results obtained in good regularisation schemes, such as
dimensional reduction, can one see that $F_{4K\zeta}$ is the most
convenient, but {\em not} more physical, starting point in
$\zeta$-function regularisation.\tnote{I think I said the
following already. Thirdly, the appearance of such extra terms in
$\zeta$-function regularisation calculations has nothing to do
with normal ordering contrary to suggestions in
\cite{EFVZ,MT98,MT}. The existence of such extra terms requires UV
infinities, some of which can be removed using normal ordering,
and in this sense the two ideas can be linked. However, as the
dimensional reduction analysis shows, normal ordering does not
create any problems or extra terms in general. It is the way that
UV divergences are regulated which is the key.}

Thirdly, as was noted at the start, the extra terms in the free
Bose gas results can sometimes be described in terms of
Multiplicative Anomalies.  Thus the results here throw some light
on the importance of Multiplicative Anomalies in general. My
results for this model confirm what has been noted elsewhere,
namely that it is essential for mathematical consistency that
Multiplicative Anomalies are included when using $\zeta$-function
regularisation. However, the interpretation given here, and in the
model used in \cite{TSEanom}, suggest that Multiplicative
Anomalies have no novel physical content. They are merely
contributions from high energy modes reflecting the way that
$\zeta$-function regularisation cuts off these modes in a physical
parameter dependent manner.

Finally, these conclusions lead me to question the extraction of
physical information from any $\zeta$-function regularisation
calculation, even after Multiplicative Anomalies are accounted
for. The analysis here was for one of the simplest examples, a
free complex scalar field on a simple flat Euclidean $T^1 \times
S^3$ space-time\footnote{The split of \tref{gencalc} and the role
of UV divergences in this problem suggests that only the short
distance behaviour is important so results will probably be
repeated in many other space-times.}. Yet I have shown that it is
almost impossible to identify the correct physics from the
$\zeta$-function regularisation results alone. It required a
comparison with dimensional regularisation to do that. However,
$\zeta$-function regularisation calculations are often used in
much more complicated situations where surely it will be even
harder to see the true physics from the UV regularisation
remnants.  I therefore believe that it will be almost impossible
to identify the true physics rather than some regularisation
remnants in many calculations using $\zeta$-function
regularisation.

\section*{Acknowledgement}

I thank E.Elizalde, A.Filippi, A.Flachi, J. McKenzie-Smith,
I.Moss, W.Naylor, R.Rivers, and D.Toms for useful discussions.


\renewcommand{\thesection}{\Alph{section}}
\setcounter{section}{0}

\section{Calculational methods.}

\subsection{Finite temperature calculations.}\label{appftcalc}

First note that doing the Euclidean energy sum can be done using
contour integration methods \cite{LvW}.  Suppose I have the
following.
\begin{eqnarray}
 F = \int_\beta dk_4\; g(k_4) &=& {1 \over \beta} \sum_{n} g(-iz = \frac{2\pi n}{\beta}).
 \label{excalc}
 \\
 &=& \frac{1}{2 \pi } \oint_C dz \; \frac{1}{2}
 \coth(\frac{\beta z}{2}) g(-iz).
 \label{excalc2}
\end{eqnarray}
The function $g$ varies in this paper but one example is
\begin{equation}\label{ftex}
 g(k_4) = -\frac{1}{s} \intdtk [(k_4+i \mu)^2 +\omegak^2]^{-s} .
\end{equation}
In this form, all the $\mu$ dependence is in the propagators and
hence in the $g$'s, and the energy sums are over pure Euclidean
discrete values \cite{TSEdal}. I use a subscript $\beta$ on
integrals as a short hand for the Matsubara sum
\begin{equation}\label{matsum}
  \int_\beta dk_4 \; g(k_4) \equiv \frac{1}{\beta} \sum_n g(k_4= 2
  \pi n \beta^{-1})
\end{equation}
The contour $C$ in \tref{excalc2} is made of several pieces, each
a small circle centred on the Matsubara frequencies $z=2 \pi i
n/\beta$, running in the positive direction \cite{LvW}. The one
complication in the types of integral I am looking at is that
there are cuts which may run along the real $z$ (Minkowski
energy) axis as the integrands are logarithms or polynomials in
energy raised to some non-integer power.  At the same time for the
Bosonic fields being considered I have a Matsubara frequency at
zero energy and therefore lying on the real axis. Thus I must
distort the cuts to run either slightly above or slightly below
the real $z$ axis at $z=0$. Normally this is trivial but the
presence of a poles in the $\coth$ function at that point
complicates matters there. Luckily, one can prove that bosonic
functions, such as those relevant to the free energy expressions I
am studying, have zero discontinuity at zero energy
\cite{TSEzegf} so the cuts may be distorted to either side without
changing the final answer.

Now having dealt with that technicality, I distort the contour,
expanding the contours round each of the poles on the imaginary
energy axis until they merge and I am left with a contour running
up the right hand side, and down the left hand side of the
imaginary energy axis, i.e.\ from $-i\infty+\epsilon$ to from
$+i\infty+\epsilon$  and then from $+i\infty-\epsilon$ to
$-i\infty-\epsilon$ ($0<\epsilon \ll 1$). Now I add on
semicircular contours running at $|z|=\infty$ in each of the
${\rm Re}(z) >0 $, ${\rm Re}(z)<0$ half-planes. The integrand
should be zero on these contours.  The $\coth$ and $\tanh$
functions are constant there and usually the form I am given for
$f(z)$ dies away at large $z$.  I am then sure that I have added
nothing to the integral by adding these semicircular pieces to the
contour.

To make the result clearer I can rewrite the $\coth$ functions
using
\begin{eqnarray}
\frac{1}{2} \coth(\frac{\beta z}{2})
 &=& +\frac{1}{2} + \frac{1}{e^{+\beta z} - 1}
 = -\frac{1}{2} - \frac{1}{e^{-\beta z} - 1} ,
\end{eqnarray}
using the first (second) expression for the ${\rm Re}(z)> (<) 0$
part of the curve.  This leaves
\begin{eqnarray}
 F &=& F_0 + F_\beta, \; \; \; \; \;
 F_0 =
 \int_{-i\infty}^{+i\infty}{-i \over 2 \pi }  dz \; g(-iz)
 \nnel
 F_\beta &=& \sum_{\{z_0\}} (-1)^{{\rm sgn}({z_0})} {\rm Res} \left[
 \left( \frac{1}{e^{\beta z*{\rm sgn}{ ( {z_0} ) } } - 1} g(-iz) \right)
 \right]
 \nonumber
 \\
 {} &&
 + {1 \over 2 \pi i} \int_{cuts} dz \; (-1)^{{\rm sgn}({z_0})}
 {\rm Disc}\left[g(-iz) \left({1 \over e^{\beta z *{\rm sgn}{ ( {z_0} ) } } -1} \right) \right]
 \label{Ip4}
\end{eqnarray}
where $\{z_0\}$ are the set of poles of $g$. ${\rm Res}[\ldots]$
and ${\rm Disc[\ldots]}$ indicate that the residues at the poles
and the discontinuities along the real $z$ cuts should be
calculated. The ${\rm sgn}$ function is given by ${\rm sgn}(z)=+1$
($-1$) if $Re(z)>1$ ($<1$).

The first term $F_0$ is the Euclidean zero temperature energy
integral.  It has no explicit temperature dependence, though $g$
contains chemical potential terms coming from the propagators.
This shows that the UV finite $T>0$ contribution $F_\beta$ can be
pulled off. All the discussion in this paper centres around terms
which are independent of temperature, though not of $\mu$ or
$m$.\footnote{I believe that one could easily make temperature
dependent Multiplicative Anomalies given the explanation for
Multiplicative Anomalies given elsewhere in this paper.  All one
needs to do is to make the UV regulating function temperature
dependent.} Thus one can just focus on the UV infinite, $T=0$
terms $F_0$ where one merely replaces the Matsubara energy sums in
the original expression by a straight Euclidean
integral.\footnote{I am assuming that one has kept the chemical
potential in the propagators and not put it in the boundary
conditions which is an alternative approach \cite{LvW,TSEdal}.}
The chemical potential remains encoded here in the quadratic
operators and propagators, present even in $T=0$ $F_0$
expressions.

\tpre{One point of warning.  A free Bose gas with non-zero charge
density {\em always} has $|\mu|=m$ at $T=0$ if $Q/V \neq 0$ since
all particles must lie in the ground state in this case \cite{Ka}.
One must always remember that $\mu$ is an implicit function of
temperature since it is the charge density $Q/V$ and not $\mu$
which is the physical observable.  For a free Bose gas it may not
always be possible to ensure $|\mu|<m$ for any given temperature
and charge density.}

\subsection{Dimensional Regularisation}\label{appdr}

In this appendix, I will summarize the most relevant aspects of
the literature and fix my notation.  References such as Collins
\cite{Co} provide a much more detailed and careful derivation of
the necessary mathematical theorems.  However, I will also comment
on how to apply this at finite temperature there is a crucial
point which has direct relevance to the discussion of
Multiplicative Anomalies.

In dimensional regulation, one adds extra components to the loop
momenta vectors in `directions' orthogonal to all other
four-vectors. The number of these `extra' dimensions is then taken
to be $-2\epsilon$, close to, but not exactly zero. More precisely
the calculations are performed at a value of $\epsilon$ where the
integrals are well defined and then analytically continued to
small finite $\epsilon$.  Luckily there is no need to do these
stages in detail as provided one follows some basic rules,
dimensional regularisation can be implemented using some standard
identities which guarantee mathematical consistency, at least for
the simple field theories considered here.

Suppose one has a single $d$-dimensional Euclidean integration
over a {\em Euclidean} loop momentum variable $k$ with no external
four vectors in the problem.   If working at zero temperature and
if the integrand depends only on $K=k^2$, then dimensional
regularisation is implemented as follows
\begin{eqnarray}
 \int \frac{d^Dk}{(2\pi)^D} \stackrel{Dim.Reg.}{\rightarrow}
 \int \frac{d^dk}{(2\pi)^d}
 &=& \int \frac{d^Dk}{(2\pi)^D} R_\epsilon{k}
 \nnel
 &=& g(\epsilon) c(d) M^{2 \epsilon} . \int dK \; K^{(d-2)/2},
 \label{drdef}
 \; \; \;
 K=k^2
 \nnel
 d &:=& D-2\epsilon,
 \; \; \;
 D \in \bbZ^+
 \nnel
 [c(d)]^{-1} &:=& {(4 \pi)^{d/2}} {\Gamma(d/2)} , \; \; \;
 g(\epsilon) = 1
 \label{cddef}
\end{eqnarray} where $M$ is the renormalisation scale.
Crudely, dimensional regulation controls the UV by inserting a
regulating function $R(k)$ of the form $R_\epsilon=
(k^2/M^2)^{-\epsilon}$, where $\epsilon \not\in \bbZ$, into the
integrand.  The UV divergences appear as $1/\epsilon$ terms. Note
that IR divergences will appear in the same terms, a draw back for
dimensional reduction in general but not an issue in the free Bose
gas model.

One useful identity in dimensional regularisation is
\bea
 \int d^pk' \int d^qk'' \; f( k'^2 +k''^2) =
   \int d^{p+q} k \; f(k^2) .
 \label{drprod}
\eea
Demanding that this identity, and others like it, are  satisfied,
fixes the forms of $c(d)$ and $g(\epsilon)$ \cite{Co}. For this to
hold it is important that $c(d)$ has the form given in
\tref{cddef}.

For more complicated integrals, e.g.\ with external loop momenta,
consistency in dimensional reduction requires that one regulates
only using components of the loop momenta orthogonal to any
external four-vectors in the problem \cite{Co}. When working in a
heat bath though, there is the velocity of the heat bath with
respect to the observer, $u^\mu$. I have chosen to work in the
rest frame of the heat bath where $u^\mu= (1,\tsevec{0})$. Thus
$k_4 \equiv k.u$ and, by the rules of dimensional regularisation,
$k_4$ can not be included in the regulating function.\tnote{Note
that at $T=0$ and $\mu=0$, where there is no dependence on
$u^\mu$, this expression is then equivalent to the usual
four-dimensional result due to \tref{drprod}.} So I have in
four-dimensions in thermal field theory for integrals with no
external momenta and only one loop variable
\begin{eqnarray}
\int_\beta \dfk  & \stackrel{Dim.Reg.}{\rightarrow}&
 \int_\beta \frac{d^{4-2\epsilon} k \; M^{-2\epsilon} }{(2\pi)^{4-2 \epsilon}}
 :=
 \int_\beta \dkf  \int \frac{d^{3-2\epsilon} \veck}{(2\pi)^{3-2\epsilon}}
 \nnel
 && = c(3-2\epsilon) M^{2 \epsilon} . \frac{1}{\beta} \sum_{n}
 \int dK \; K^{1/2-\epsilon}
 ,
\label{drftdef}
\\
&& k_4 := \frac{2\pi n}{\beta}, \; \; \; n \in \bbZ \label{k4def}
\end{eqnarray}
and $c(d)$ is given in \tref{cddef}.  One can perform the energy
sum in the usual manner assured that the result is finite even
after the spatial integration provided one keeps $\epsilon$
non-integer. One can then separate off the zero temperature and
finite temperature parts as discussed in appendix \ref{appftcalc}
and shown in \tref{Ip4}. The former can then be returned to the
usual zero-temperature $(4-2\epsilon)$-dimensional form through
the use of \tref{drprod}. The finite temperature part can have the
regulator removed, $\epsilon=0$, since it is UV finite.

It is absolutely crucial that no regulation occurs in the energy
variable $k_4=k.u$.  So while the replacement
\begin{eqnarray}
 \int_\beta \frac{d^4k}{(2\pi)^4}
 & {\rightarrow}&
 \frac{1}{\beta} \sum_{n \in \bbZ } \left(\frac{k_4^2}{M^2}\right)^{-\epsilon}
 \int \frac{d^{3} k}{(2\pi)^{3}}, \; \; \; (k_4 = 2\pi n /\beta)
 \label{ddrftdef}
\end{eqnarray}
may also regulate the divergent integrals, the resulting
expressions will not in general be mathematically consistent.  In
other words, this type of regulation may have Multiplicative
Anomalies!\tnote{Since the $T>0$ corrections are all UV finite, in
many cases there will not be a problem.  However, if there are
chemical potential terms present, these are present even in $T=0$
expressions with their UV divergences and they shift the energies
in the integrands.  Thus I am again mixing physical parameters in
the regulating scheme and this is when Multiplicative Anomalies
will appear. {The same problem would occur in most regulating
schemes if I apply them to both energy and spatial momenta.  For
instance a simple momentum cutoff should be enforced only on the
spatial momenta and not on the whole Euclidean four-vector space
as is normally done at $T=0$.} {Lattices? Or since they are in
coordinate space are they safe?} {Show this!!!}}

\tpre{Finally, note that there is a close relationship between the
family of different regularisation schemes and the families of
renormalisation schemes.  In dimensional reduction we can always
change the definition of regularised integrals by multiplying by
$g(\epsilon)$, where $g(z)$ is a function which is unit valued and
analytic at $z=0$.  However, if one was to implement MS
renormalisation but with a $g$ dependent regularisation, i.e.\
just subtract the leading $1/\epsilon$ poles whatever $g$ is, we
would get extra finite terms proportional to $g'(0)$. If $g$ is a
simple algebraic function, and has no additional dependence on
physical variables, then these can be absorbed in constant
redefinitions of the renormalisation scale $M$.  For instance, in
dimensional reduction it is common to use $\overline{{\rm MS}}$
(Modified Minimal Subtraction scheme) which is equivalent to MS
with a $g(z) = 1+ z(\ln(4\pi) - \gamma) + O(z^2)$\tnote{???} and
the relationship between these two schemes is simple.
Alternatively, the effects of $g(\epsilon)$ can be encoded through
a conformal transformation on the dimensionless parameter
$\epsilon$.}

\subsection{$\zeta$-function regularisation}\label{appzfr}

I implement $\zeta$-function regularisation scheme
\cite{DC,Ball,EORBZ} as follows
\bea
 \Tr \ln \{ A \}
 & \stackrel{ \zeta -func.reg. }{ \rightarrow } &
 g\left(\frac{2s}{a}\right) \frac{a}{2s}
 \Tr \left\{ \left( \frac{A}{M^{a}} \right)^{2s/a} \right\}
 = \frac{a}{s} \zeta \left( \frac{2s}{a} | \frac{A}{M} \right)
 g\left(\frac{2s}{a}\right)
 \label{zfrlnop}
 \\
 \Tr \{ A \}
 & \stackrel{ \zeta-func.reg.}{\rightarrow} &
 g\left(\frac{2s}{a}\right) \Tr \left\{ \left( \frac{A}{M^{a}} \right)^{1+2s/a} \right\}
 = \zeta \left( 1 + \frac{2s}{a} | \frac{A}{M} \right)
 g\left(\frac{2s}{a}\right)
 \label{zfrgenop}
\eea
where $a \in \bbZ^+$ is both the order and dimension of the
operator $A$, i.e.\ I assume that $A \sim |k|^a$ in the UV limit.
The function $g$ is set to one in standard calculations and I will
do that here, though see below for further comments about this.
Roughly speaking, regulation is achieved by
\begin{equation}
 \int \frac{d^Dk}{(2\pi)^D} A {\rightarrow} \int \frac{d^Dk}{(2\pi)^D}
 R_\zeta A
\end{equation}
where $R_\zeta = [A/M^a]^{-2s/a}$. Thus both dimensional reduction
and $\zeta$-function regularisation use non-integer powers,
$\epsilon$ or $s$, of $k$ to achieve their regulation. The key
difference for this work is the fact that $\zeta$-function
regularisation's regulating function involves the operator $A$,
and consequently varies with changes in any physical parameters in
$A$.  The regulating function of dimensional reduction does not
have this dependence. This ensures that dimensional
regularisation satisfies certain basic identities \cite{Co}, while
$\zeta$-function regularisation does not as appendix \ref{appma}
notes. In fact this is the essence of the Multiplicative Anomalies
in $\zeta$-function regularisation, while a great deal of work has
been done to ensure mathematical consistency of basic algebraic
identities such as \tref{drprod} in dimensional regularisation
\cite{Co}. A few further points are worth making.

Firstly I have ensured that naively the expressions always have an
overall factor of $(M^2)^{-s}$ to mimic the $(M^2)^{-s}$  factor
in the dimensional regularisation expressions.  This aids
comparison of the results obtained using the two regularisation
schemes.  More importantly though, it means that I have made a
very specific choice for the form of the regulating power, which
is not simply a small parameter $s$ but a specific multiple of it.
This is to ensure that the poles will cancel when calculating
different terms in Multiplicative Anomalies as shown in
\tref{anomex}, i.e.\ I can use the same $s$ parameter in all
cases. In the same way, when comparing $\zeta$-function
regularisation and dimensional regularisation expressions, a
simple equality $s=\epsilon$ will be sufficient to match such
expressions.  This is to be contrasted with the renomalisation
scale $M$ which will often be rescaled when comparing different
expressions.

Secondly, I choose to mimic dimensional regularisation in another
way by ensuring that the UV divergence always appears as a
$s^{-1}$ pole with the physics contained in the $O(s^0)$ term of a
small $s$ expansion.  The generalised $\zeta$-function
$\zeta(z|A)$ (for relevant operators $A$) is finite at $z=0$ but
in general has $s^{-1}$ poles at $z=n+s, n=1,2,...$.  Thus
creating $s^{-1}$ pole is only an issue for the regularisation of
logarithm of operators where I have had to add an overall factor
of $\frac{a}{2s}$ in \tref{zfrlnop}. In fact this is very natural
as the following example will show.

Consider
\beq
H= \ln(AB)- \ln(A) - \ln (B) =0 \label{anomex}
\eeq
where $A$ and $B$ could be ordinary numbers or more complicated
objects. These logarithms can be represented as the second term in
the following series
\beq
[(A)^{s/a}] = 1 + \frac{s}{a} \ln(A) + \frac{s^2}{2a^2} (\ln(A))^2
+O(s^3)
\eeq
The parameter $a$ has no special meaning here but one can add it
in anticipation of its role in the field theory case where $A
\sim |k^a|$ as $|k| \rightarrow \infty$. Therefore I can
construct a representation of $H$ through the replacement
\beq
 \ln (A)  \rightarrow \frac{a}{s} [(A)^{s/a}]
 \label{Areg}
\eeq
where $s$ is small. Similarly for $\ln(B)$ I use ${s/b}$ rather
than ${s/a}$ where $b$ is arbitrary in this simple context. I must
then use
\beq
 \ln (AB) \rightarrow \frac{a+b}{s} [(AB)^{s/(a+b)}]
 \label{ABreg}
\eeq
to ensure that the $1/s$ infinities cancel and I indeed find that
\bea
 H \rightarrow H_s &=&
 - \frac{s}{2ab(a+b)} \left[ \ln \left( \frac{A^b}{B^a} \right)
 \right]^2
 + O(s^2)
 \label{anomexres}
\eea
which is zero as $s \rightarrow 0$.  The first $O(1/s)$ terms in
the expression cancel because of the judicious choice of
regularisation parameters for \tref{Areg} and \tref{ABreg}. The
next $O(s^0)$ terms are just the original simple logarithms which,
if we have constructed a sensible representation of the original
expression \tref{anomex}, must cancel. Thus the first non-zero
terms are the $O(s^1)$ terms but, provided all the terms are
finite, this term will be zero in the limit $s=0$ where we expect
to recover the expression $H$.  However, if one is taking a trace
over some variable upon which $A$ and $B$ depend, one can imagine
that the trace may be divergent even if $A$ and $B$ are not. This
may correspond to an additional $1/s$ pole being generated which
can then combine with these $O(s^1)$ terms to give a non-zero
contribution in the $s \rightarrow 0$ limit, i.e.\ an anomalous $H
\neq 0$ result. Note the close similarity between the form of the
expression for $H_s$ and the Wodzkicki formula \tref{wresanom} for
the logarithmic Multiplicative Anomaly anomaly.

One often hears that one of the advantages of $\zeta$-function
regularisation is that it is finite.  However all that is meant is
that $\zeta(z|A)$ is finite near $z=0$ so one may define the $\Tr
\ln A = \ln \det A = \zeta'(z=0|A)$.  In the context, of QFT this
is not especially amazing. The physics is still in the second term
of some expansion and one is merely giving a prescription for
extracting this term. In my implementation of $\zeta$-function
regularisation, I have merely multiplied by a factor of $1/s$ when
regulating $\Tr\ln$ expressions in order to keep other
$\zeta$-function regularisation expressions and indeed
dimensional reduction expressions on a similar footing. Thus the
usual definition of $\zeta$-function regularisation, as applied to
vacuum energy densities, corresponds to dropping what are the
leading $1/s$ poles in my expressions, i.e.\ traditional
$\zeta$-function {\em regularisation} is also a `minimal
subtraction' {\em renormalisation} scheme. Ignoring the first term
in the small $s$ expansion is merely a particular renormalisation
scheme, not just a simple regularisation\footnote{Regularisation
is process where infinities are turned into large but finite
terms. Renormalisation is a process which expresses bare
parameters in terms of finite values which can be related to
physical measurements.  The latter must always be performed in
QFT, even if the theory is finite, as quantum fluctuations are
always present. It will though remove any infinities present in
expressions for physical quantities.}. Put another way one can
mimic the usual implementation of $\zeta$-function regularisation
for vacuum energy densities in dimensional regularisation. To do
this the usual dimensional regularisation expression is multiplied
by $\epsilon$, which makes the expression finite in the same way
that $\zeta(0|A)$ is. One would then define vacuum energies $\Tr
\ln A = \ln \det A$ to be the $\epsilon$ derivative of that
combination, i.e.\ I take just the second term in a small epsilon
expansion and drop the first term. This is of course just the
Minimal Subtraction renormalisation scheme.\tnote{Thus by analogy,
$\zeta$-function regularistion as normally applied to vacuum
energy terms is actually a renormalisation and it is the Minimal
Subtraction scheme of the family of Schwinger proper-time
regularisations \cite{Ball}.}

Finally, one may always multiply the $\zeta$-function
regularisation expressions, \tref{zfrlnop} and \tref{zfrgenop}, by
any function $g(2s/a)$ provided this function $g(z)$ is analytic
and of value one at $z=0$.  Its introduction merely moves us
through the family Schwinger proper-time regularisations
\cite{Ball}, changing the finite terms by an amount proportional
to $g'(0)$. The $g(0)=1$ condition can be imposed with out loss of
generality as any other value is equivalent to a rescaling of the
small regulating $s$ parameter.  This freedom is used to ensure
that poles cancel in Multiplicative Anomalies, as discussed for
equation \tref{Areg} and \tref{ABreg}.  One example of interest is
$\bar{g}=g(2s/a) = s\Gamma(s)$ for $\Tr \ln$ problems so that from
\tref{zfrlnop}
\begin{equation}\label{SPTDRdef}
  \Tr \ln A \stackrel{\zeta-f.reg.\ \bar{g}}{\rightarrow}
  \zeta \left( \frac{2s}{a} | \frac{A}{M} \right) \Gamma(s) ,
\end{equation}
I will refer to this as SPTDR (Schwinger Proper Time Dimensional
Regularisation). This member of the family of Schwinger proper
time regularisation schemes is sometimes referred to as plain
`dimensional regularisation' in some of the literature on
$\zeta$-function methods, as in\tnote{Equation (1.6) in
\cite{CZ}.} \cite{CZ}.  However SPTDR is {\em not} the dimensional
regularisation of the particle physics literature such as
\cite{PDG,Co} and as described in appendix \ref{appdr}.  The
review of Ball \cite{Ball} makes this distinction clear and simple
examples confirm this view.

\subsection{The Wodzicki residue.}\label{appwod}

In the theory of elliptic pseudo-differential operators, there is
a unique extension of the Diximier trace to elliptic
pseudo-differential operators and Wodzicki gave the explicit form
\cite{Wo,Wo2}. This beautiful result of modern mathematics is
central to several areas, such as Non-Commutative Geometry (e.g.\
see \cite{Connes}). However, there is a simple formula for
calculating it and a simple relation between the Multiplicative
Anomaly of \tref{adef} and the Wodzicki residue.

The simple recipe to find the Wodzicki residue\tprenote{In
\cite{Wo} see sec.\ 7.13, pp.176. In \cite{Kass} see section 6.5,
pp.225-226, especially his eqn.(9). Also note the comments in
section 1.2, eqn.(4) which cites Wodzicki's 1984 thesis for a
residue formula. See also Wodzicki \cite{Wo2}.} is as follows
\cite{Wo,Wo2,Kass}.  The ``complete symbol'' for our operators is
$A(x,k) := e^{-ikx} A e^{ikx}$. Extract the UV behaviour as an
asymptotic expansion
\begin{equation}\label{asym}
A(x,tk) = \sum_{j=a}^{-\infty} t^j A_j(x,k)
\end{equation}
where $a$ is the order of the operator.  The Wodzicki residue $W$
of an operator $A$ is then
\begin{equation}\label{Wres}
{\rm res}_W (A) = \int_{\cal M} d^Dx \int \dslash^Dk M^2
\deltaslash (k^2-M^2) A_{-D}(x,k)
\end{equation}
Note that in many places (all the ones I found) the regularisation
scale $M$ is set to one, that is the calculations are done with
all dimensionful parameters being measured in units of $M$.  While
perfectly acceptable mathematically, this does mean that the
renomalisation scale is hidden even though it is extremely
important when extracting real physical numbers from the
mathematical results \cite{PDG}.

There is a simple link to the Cauchy residues of $\zeta$-functions
through
\begin{equation}\label{wreszeta}
{\rm Res}_{s=\sigma} [ \zeta(s|A)] = a^{-1} {{\rm res}_W
(A^{-\sigma}) }
\end{equation}
and finally to the Multiplicative Anomaly through
\begin{equation}\label{wresanom}
{a_\zeta(A,B)} = (2ab(a+b))^{-1} {{\rm res}_W \left[ \{ \ln ( A^b
B^{-a} ) \}^2 \right] }
\end{equation}
Note the strange expressions in \tref{wresanom} are easily
obtained in a simple analysis of logarithms of numbers which gives
some insight into this expression, see \tref{anomexres}.

\subsubsection{Conjecture for a generalised
Multiplicative Anomaly formula}\label{appwodconj}

The Wodzicki residue formula \tref{wresanom} is for $\Tr \ln$
expressions \tref{adef}, relevant to zero point energy
calculations in general and to all the four-dimensional examples
considered here. However simple traces over other operators can
give rise to Multiplicative Anomalies, as the examples
\tref{anomexp}, \tref{anomcshift} and \tref{anomcmult} show. One
can try to write these examples in terms of $\tr \{ z \} = \tr \ln
\{ A \}$, with $A = \exp \{ z \}$, and then try to use the
Wodzicki residue formula for the Multiplicative Anomaly.
Unfortunately, the UV behaviour of the resulting operators $A$ is
then exponentially divergent or suppressed and the formula does
not work. However, since direct calculations in $\zeta$-function
regularisation in these cases does show the existence of an
Multiplicative Anomaly, one wonders if there is not a formula
similar to the one based on the Wodzicki residue.  I have made the
following conjecture. Suppose a Multiplicative Anomaly is given by
an unregulated expression $\calF$
\beq
 \calF = \sum_j \int \dslash^Dk \; F_j(k) ,
 \eeq
 where
 \beq
 \sum_j  F_j(k) = 0 \; \; \forall |k| < \infty
 \label{constraint}
\eeq
The operators $F_j$ in each term must have the same
dimension which, without loss of generality, I will take to be the
same as the order $a$, that is $F\sim |k|^a$ as $k \rightarrow
\infty$. Let me now write down the $\zeta$-function regularisation
form
\beq
\calF_\zeta = \sum_j \int \dslash^Dk \; [F_j(k)]^{1+s/a} M^{-s}
\eeq
where for definiteness I choose to regulate the expression so that
the UV behaviour is $|k|^{a+s}$ and the renormalisation scale
always appears as the simple $M^{-s}$ factor. Now suppose I make a
double expansion of $F^{1+s/a}$ in s and $1/k$ and define
\bea
[F_j(t \kbar)]^{1+s/a} &=&
 |k|^{a+s} \sum_{m=0}^\infty \sum_{n=0}^\infty s^m t^{-n}
  \frac{1}{m!n!}
 \tilde{f}_{j,m,n}
 (\kbar).
\eea
The interest is in the UV divergences so it makes sense to
separate off the integral over the $|k|$ through
\bea
1 &=& \int_0^\infty dt \; 2 t \delta(t^2 - k^2/M^2) .
\eea
By looking at the
formula for the $\Tr \ln$ examples and trying a few examples, I
have made the following conjecture
\bea
 \calF_\zeta & \propto &
 \sum_{j} 2 \int \dslash^D\kbar  \; \delta(1- \kbar^2/M^2)
 M^{a} .
 \frac{1}{(D+a)!}
 \tilde{f}_{j,2,D+a} (\kbar) .
 \label{anomconj}
\eea
i.e.\ only the $O(s^2)$ and $|k|^{-D-a}$ term of the
$f_j^{1+s/a}$ expansion contributes. The reason for trying the
$n=D+a$ term comes from the original Multiplicative Anomaly
formula using Wodzicki residue.  There the term which appears is
the one in the asymptotic expansion of the operator which is right
on the boundary between divergence and convergence, i.e.\ the
logarithmic divergent term which in this case is the $n=D+a$ term.

The reason for trying $m=2$ term in the conjectured form
\tref{anomconj} is that it is quickly apparent that the $m=0$ and
$m=1$ terms are zero. That this must be so follows from the two
constraints on Multiplicative Anomaly expressions. Firstly it
should not itself be divergent so $1/s$ terms should cancel when
doing the sum over $j$. In general one must choose the $s$
parameter used to regulate each term in the $j$ sum in just the
right way to achieve this, and the definition used here ensures
this, as noted in appendix \ref{appzfr}. Secondly, the
Multiplicative Anomaly expressions are naively zero in the first
place \tref{constraint}, and this is linked to the fact that the
sum over $j$ of the $O(s^0)$ terms must be zero. Thus it can only
be the third term in the $s$ series of $f$ that is relevant,
presumably because in a proper calculation it is mixing with
divergent terms. See \tref{anomex} for a simple example of this.


\section{A collection of Multiplicative Anomalies}\label{appma}

In \cite{TSEanom} it was suggested that Multiplicative Anomalies
appear only if three criteria hold, namely
\begin{enumerate}
\item the terms in an Multiplicative Anomaly must contain infinities,
\item the regularisation scheme used must mix physical parameters
(mass, chemical potential, etc.) in the artificial function used
to cutoff UV momenta, and
\item the different terms in the Multiplicative Anomaly must contain
different combinations of these physical parameters.
\end{enumerate}
Thus regularisation schemes such as dimensional regularisation,
physical cutoffs (e.g.\ lattice) and ones typically used in
particle physics never have Multiplicative Anomalies, as the
examples in \cite{TSEanom} show.

However, $\zeta$-function regularisation and all other members of
the family of Schwinger proper time regularisations \cite{Ball}
are {plagued} by Multiplicative Anomalies.  In general with these
regularisation schemes {\em regulated expressions will fail to
obey many algebraic identities naively satisfied by their
unregulated counterparts}.  Is there any identity which
$\zeta$-function regularisation respects?

The criteria above are merely necessary, not sufficient.  For
instance the simplest example of an Multiplicative Anomaly in QFT
found in \cite{EVZ} is non-zero only in even dimensions from four
upwards. However, numerous examples of non-zero Multiplicative
Anomalies are now known. As some are relevant for the discussion
of the free Bose gas, I will give a brief list. Here $A,B$ are
suitable operators, $\alpha$ is a c-number.

The original example
\begin{equation}\label{anomze}
\Tr \ln(AB) \neq \Tr \ln(A) + \Tr \ln(B)
\end{equation}
was given by Elizalde, Vanzo and Zerbini for several simple cases
in field theory, including two real scalar fields of different
masses in flat Euclidean space-time \cite{EVZ}.  Physically it is
important when considering the vacuum energy of two free scalar
fields as in cosmology or in the Casimir effect.  It also appears
in the free Bose gas case, in particular the four-dimensional
expressions \tref{F4Adef}, \tref{F4Kdef} and \tref{F4Ldef}, which
were first given in \cite{EFVZ}.

The problem of Multiplicative Anomalies appears with Green
functions as well as with vacuum energy densities considered in
\cite{EVZ}.  Consider
\begin{equation}\label{anomexp}
  \Tr \{ \frac{1}{A(A+\alpha)}\} \neq \left(\Tr\{\frac{1}{A} \} -
\Tr\{ \frac{1}{A+\alpha} \}\right)\frac{1}{\alpha}
\end{equation}
It is formally related to the first Multiplicative Anomaly, as
defined in \tref{adef}, by the replacement in \tref{adef} $A
\rightarrow \exp \{ 1/A \}$ and $B \rightarrow \exp \{
1/(A+\alpha) \}$.  Examples of this failing with $\zeta$-function
regularisation were shown in \cite{TSEanom} for two real scalar
fields of different masses\footnote{The analysis in \cite{EFVZ}
confirms that of \cite{TSEanom} contrary to what is stated in
\cite{EFVZ3}}\tnote{The regularisation used around equation (4) in
\cite{EFVZ3} is in fact the usual dimensional regularisation of
particle physics.} with $A=\Delta^{-1}, \alpha = \delta m^2$. The
Multiplicative Anomaly is then related to expectation values of
the squares of the two real fields, $\langle (\phi_1^2 - \phi_2^2)
\rangle \neq \langle \phi_1^2 \rangle - \langle \phi_2^2 \rangle$.

Mass shifts are perhaps the most simple interaction possible and
often appear in QFT.  They are also afflicted by an Multiplicative
Anomaly in $\zeta$-function regularisation, for instance
\begin{equation}\label{anommshift}
\Tr \{\ln(A+\alpha)\} \neq \Tr \{\ln(A)\} - \sum_{n=1}^\infty
\frac{(-\alpha)^n}{n} \Tr \{A^{-n} \} .
\end{equation}
An example of this Multiplicative Anomaly was given in
\cite{TSEanom} for a single real scalar field, where one looks at
a mass shift of the scalar field, $A=\Delta^{-1}, \alpha = \delta
m^2$

A generalisation of the Multiplicative Anomaly in \tref{anomze} is
\begin{equation}\label{anomdimred}
  \Tr \{ \ln(\prod_n A_n ) \} \neq \sum_n \Tr \{\ln(A_n)\} .
\end{equation}
This version is relevant in the context of dimensional reduction
\cite{FSZ,Su,CZ}.

Simple shifts by c-numbers lead to Multiplicative Anomalies
\begin{equation}\label{anomcshift}
 \Tr \{ A + \alpha\} + \Tr \{ A - \alpha\} \neq  \Tr \{ 2A \} ,
\end{equation}
showing that the linear property of the trace is not preserved in
$\zeta$-function regularisation {\it pace} \cite{EFVZ3}.  The
first example of a Multiplicative Anomaly of this type was first
noted by McKenzie-Smith and Toms \cite{MT98,MT} in the context of
the free Bose gas using the three-dimensional expressions of
\tref{F3mudef} and \tref{F3def}. Again if in \tref{adef} I make
the replacements $A \rightarrow \exp\{ -A - \alpha\}$ and $B
\rightarrow \exp\{ -A + \alpha\}$ this links this Multiplicative
Anomaly to the original one.

Finally, if one needed any further convincing about the existence
of Multiplicative Anomalies, one need look no further than simple
c-number multiplication of operators as in $\zeta$-function
regularisation\tnote{This is sometimes called the trace anomaly.}
\begin{equation}\label{anomcmult}
\Tr \{ \alpha A\}  \neq \alpha \Tr \{ A \} .
\end{equation}
It is needed as well as \tref{anomcshift} when relating $F_3$ and
$F_{3\mu}$ It is easy to see that if $\Tr \{ A \}$ is regularised
using $\zeta$-function regularisation and becomes $\Tr \{ A^{1+s}
\} = a_{-1} s^{-1} + a_0 + O(s)$, then there is a Multiplicative
Anomaly of
\begin{equation}\label{shiftanom}
a_{{\rm rescale}}({A},{\alpha}):= \Tr \{ \alpha A\}  - \alpha \Tr
\{ A \} = \alpha \ln ( \alpha ) a_{-1} .
\end{equation}
This is non-zero whenever $\Tr \{ A \}$ is infinite, just as
required by the criteria in \cite{TSEanom}.  This is no more than
a changing of the regularisation scale which always appears in QFT
e.g.\ let $A= \Delta$ the propagator for a real scalar field, one
will have to remember that I need to introduce a new scale, say
$M$, to keep the dimensions of the regularised expression correct,
e.g.\ as in \tref{F4Aedef} and \tref{F4Kzdef}. Note that this
tells us that `trivial' operations such as pulling out the factor
of two on the right hand side of \tref{anomcshift} or pulling a
minus sign outside the trace are in fact highly non trivial if
Multiplicative Anomalies are non-zero. Yet such operations are
performed frequently in formal derivations, and in particular they
abound in our canonical derivations of the free energy of the free
Bose gas in the last section.  Thus the Multiplicative Anomaly,
$a_{\rm shift}$ of \tref{ashiftdef}, appearing used in the
three-dimensional forms for the free energy encodes a failure in a
combination of \tref{anomcshift} and in \tref{anomcmult}.

The point of giving this summary has been to show how many
different algebraic operations fail when using $\zeta$-function
regularisation, even in the simplest of QFT problems involving
free fields in flat Euclidean space times.   One can make these
algebraic operations but {\em only} if one remembers to include
the Multiplicative Anomaly terms as well. It is extremely easy to
neglect these Multiplicative Anomalies but this will lead to
mathematical inconsistencies.  In this sense Multiplicative
Anomalies are of vital importance to problems using
$\zeta$-function regularisation.

Several of these Multiplicative Anomalies are relevant to the
analysis of the free energy of the free Bose gas. In particular,
the regularisation scale, here called $M$, is often shifted in the
analysis above.  The simple example \tref{shiftanom} can be
interpreted in this way.  This all gives further weight to the
idea in \cite{TSEanom} that one can take account of all
Multiplicative Anomalies by regarding them as shifts in
renormalisation scales.


\tpre{
\section{Useful formulae}

\begin{equation}\label{binexp}
  (1+x)^{\epsilon} = \sum_{n=0}^{\infty}
  \frac{\Gamma(1+\epsilon)}{\Gamma(1+n)\Gamma(1+\epsilon-n)} x^n
\end{equation}

\begin{eqnarray}
\int_0^\infty dx \; \frac{x^{\mu-1}}{(1+\beta x^p)^\nu} & =&
\frac{\beta^{-\mu/p}}{p} \frac{\Gamma(\mu/p) \Gamma(\nu -
\mu/p)}{\Gamma(\nu)} \label{GR3.251.11app}
\end{eqnarray}

\bea
\Gamma(n+\epsilon) &=& \Gamma ( n) \left( 1+ \epsilon \psi(n)
\right) , \; \; \; n \in \bbZ^+ \label{Gamnp}
\\
\Gamma(-n+\epsilon) &=& \frac{1}{\epsilon}
\frac{(-1)^{n}}{\Gamma(1+n)} \left( 1+ \epsilon \psi(1+n) \right)
, \; \; \; n \in \{ 0, \bbZ^+ \} \label{Gamnn}
\\
 \Gamma(1/2)&=& \sqrt{\pi}
 \label{Gamm1o2}
 \\
 \Gamma(3/2)&=& \half \Gamma(\half) = \frac{\sqrt{\pi}}{2}
 \label{Gam3o2}
 \\
 \Gamma(-1/2)&=& -2 \Gamma(\half) = -2 \sqrt{\pi}
 \label{Gammm1o2}
\eea

\begin{eqnarray}
 \psi(z) &:=& \frac{d \ln( \Gamma ( z) ) } {dz}
 \label{psidef}
 \\
 \psi(1+x) &=& \psi(x) + \frac{1}{x}
 \label{psix}
 \\
 \psi(1) &=& -\gamma \approx - 0.577 215 664 90 , \; \; \; \gamma
 \mbox{ is Euler's constant},
 \label{psi1}
 \\
 \psi(1/2) &=& \psi(1) - 2 \ln ( 2 )
 \label{psihalf}
 \end{eqnarray}

\begin{eqnarray}
 \int \dslash^D K &=& \frac{c(D-1)}{2\pi} \int_0^\infty dK K^{D-1}
 \int_0^\pi d\theta (sin(\theta))^{D-2}
 \\
 & =& c(D) \int_0^\infty dK K^{D-1} , \\
 && K^i = (\veck, k_d)^2, \; k_d=K \cos(\theta), \; |\veck| = K
\sin(\theta)
 \\
 c(d) &=& \left((4\pi)^{d/2} \Gamma(d/2)\right)^{-1}
\end{eqnarray}

}

\end{document}